\documentclass[12pt]{article} 
\usepackage{epsfig} 
\usepackage{citesort} 
\usepackage{times} 

\newcommand{\eq}{\begin{equation}} 
\newcommand{\eqx}{\end{equation}} 
\newcommand{\eqn}{\begin{eqnarray}} 
\newcommand{\eqnx}{\end{eqnarray}}

\newcommand{\Th}{\theta} 
\newcommand{\om}{\omega}

\newcommand{\nin}{\noindent} 
\newcommand{\lam}{\lambda} 
 
\newcommand{\seld}{\frac{d\sigma_{el}}{d\Th}(E)} 
 
\newcommand{\eps}{\epsilon} 

\begin{document} 
\nin 
\centerline{\Large DRAFT} 
\begin{center} 
{\Large \bf SPACE-TIME EVOLUTION OF ELECTRON CASCADES IN DIAMOND}\\ 
\vspace{10mm} 

{\Large Beata~Ziaja $^{\ast,\,\dag,\,\ddag}$}, 
{\Large Abraham~Sz\"{o}ke $^{\ast,\,\S}$}, 
{\Large David~van der Spoel $^{\ast}$ and} 
{\Large Janos~Hajdu $^{\ast}$} \footnote{e-mail:ziaja@tsl.uu.se, 
~szoke1@llnl.gov,~spoel@xray.bmc.uu.se,~hajdu@xray.bmc.uu.se}\\ 

{\footnotesize 
\vspace{3mm} 
$^{\ast}$ \it Department of Biochemistry, Biomedical Centre, 
\it Box 576, Uppsala University, S-75123 Uppsala, Sweden\\ 
\vspace{3mm} 

$^{\dag}$ \it Department of Theoretical Physics, 
	 Institute of Nuclear Physics, 
\it Radzikowskiego 152, 31-342 Cracow, Poland\\ 
\vspace{3mm} 
$^{\ddag}$ \it High Energy Physics, Uppsala University, 
	 P.O. Box 535, S-75121 Uppsala, Sweden 
\vspace{3mm} 

$^{\S}$ \it Lawrence Livermore National Laboratory, Livermore, 
	 CA 94551, USA\\ 
} 
\end{center} 

\vspace{5mm} 
\nin 
{\bf Corresponding author:}\\ 
Janos Hajdu, Department of Biochemistry, Biomedical Centre,\\ 
Box 576, Uppsala University, S-75123 Uppsala, Sweden\\ 
Tel:+4618 4714449, Fax:+4618 511755, E-mail:hajdu@xray.bmc.uu.se \\ \\ 
\nin 

{\bf Abstract:} 
{\footnotesize 

The impact of a primary electron initiates a cascade of secondary
electrons in solids, and these cascades play a significant role in the
dynamics of ionization. Here we describe model calculations to follow the
spatio-temporal evolution of secondary electron cascades in diamond. The
band structure of the insulator has been explicitly incorporated into the
calculations as it affects ionizations from the valence band. A
Monte-Carlo model was constructed to describe the path of electrons
following the impact of a single electron of energy $E\sim 250$ eV. This
energy is similar to the energy of an Auger electron from carbon. Two
limiting cases were considered: the case in which electrons transmit
energy to the lattice, and the case where no such energy transfer is
permitted. The results show the evolution of the secondary electron
cascades in terms of the number of electrons liberated, the spatial
distribution of these electrons, and the energy distribution among the
electrons as a function of time. The predicted ionization rates ($\sim
5$-$13$ electrons in $100$ fs) lie within the limits given by experiments
and phenomenological models. Calculation of the local electron density and
the corresponding Debye length shows that the latter is systematically
larger than the radius of the electron cloud, and it increases
exponentially with the radial size of the cascade. This means that the
long-range Coulomb field is not shielded within this cloud, and the
electron gas generated does not represent a plasma in a single impact
cascade triggered by an electron of $E\sim 250$ eV energy. This is
important as it justifies the independent-electron approximation used in
the model. At $1$ fs, the (average) spatial distribution of secondary
electrons is anisotropic with the electron cloud elongated in the
direction of the primary impact. The maximal radius of the cascade is
about $50$ \AA$\,$ at this time. At $10$ fs the cascade has a maximal radius
of $\sim 70$ \AA, and is already dominated by low energy electrons
($>50$\%, $E<10$ eV). These electrons do not contribute to ionization but
exchange energies with the lattice. As the system cools, energy is
distributed more equally, and the spatial distribution of the electron
cloud becomes isotropic. At $90$ fs, the maximal radius is about $150$
\AA. An analysis of the ionization fraction shows that ionization level
needed to create an Auger electron plasma in diamond will be reached with
a dose of $\sim 2\times 10^5$ impact X-ray photons per \AA$^2$ if these
photons arrive before the cascade electrons recombine. The Monte-Carlo
model described here could be adopted for the investigation of radiation
damage in other insulators and has implications for planned experiments
with intense femtosecond X-ray sources.  
} 
\vspace{6mm} 


The treatment of electron cascades in this paper is restricted to low
energy electrons ($E_{impact} = 250$ eV) and weakly ionized samples where only
a few interatomic bonds are perturbed. We chose
diamond as a model compound because of its significance in X-ray optics
and its relevance in modeling covalent carbon structures. The calculations
presented here can be extended to other insulators. 

In our model ionization proceeds via excitations of secondary electrons from the valence band. In all insulators the valence band is fully occupied, and any electron released from the valence band leaves a hole in the band, which becomes an independent charge carrier. In contrast, the band structure of metals is open, and its behaviour is usually well described by a (modified) free-electron-gas approximation \cite{l6}. 

Low energy electrons may undergo elastic and inelastic collisions with atoms in a solid. Since the corresponding electron wavelength is comparable with atomic dimensions and interatomic distances in solids, multiple scatterings \cite{l8} on neighbouring atoms have to be considered with low-energy electrons and calculated quantum-mechanically (QM). The QM exchange terms have to be incorporated into the interaction potential. Calculations of  elastic scattering amplitudes can be done reasonably well in the muffin-tin potential approximation \cite{l6prim,l6bis,l12}. In contrast, a fully rigorous method for including 
inelastic scattering has not yet been published. 

The main aim of this paper is to provide a more accurate description of secondary electron emission rates in solids than the currently existing models
\cite{l29,l30,models1,models2,models3,models4,akker}. This is important for a better understanding of radiation damage. 

There is a class of processes, which do not contribute to ionization. These processes affect mostly the transport of very low energy electrons ($E\ge 10$ eV), propagating through an insulator. Electrons with energies below the secondary ionization threshold undergo elastic and inelastic collisions due to  atomic vibrations, impurities or atomic vacancies \cite{secon5}. In metals, inelastic electron-electron interactions are predominant but this is not the case with insulators. In insulators, lack of electrons in the conduction band implies that the predominant energy transfer is via atomic vibrations (phonons) for very low energy electrons. These processes do not lead to additional emission of secondary electrons, since the energy gains and losses due to the phonon coupling are very small, despite the fact that such couplings happen at a rate of about $10^{14}$ phonons per second. We neglect these very low energy couplings, since they do not contribute to ionization and their influence on the energy transfer is not significant in the first $10$ fs. The model presented here predicts the total number of electrons released in the cascade correctly but misestimates the average energy of electrons in the range of $10$ to $100$ fs. For the same reason, we do not follow the evolution of cascades after $\approx 100$ fs. 

The present study goes beyond the free-electron-gas model used in our previous,
zero-dimensional model \cite{ziaja}, and incorporates a detailed description of the band structure of diamond. It also considers dispersion relations for carriers in the valence band and in the conduction band. In what follows, we first describe the MC model, and show results of $2000$ simulations from different cascades. These yield the characteristics of the secondary electron cloud as a function of time. Finally, we discuss the results in the context of their implications  for radiation damage.

\subsection*{The model:}

Calculation of elastic scattering amplitudes and angular distributions was done in the muffin-tin potential approximation by the partial wave expansion technique \cite{l6prim,l6bis,ziaja}. To perform these calculations we used programs from the Barbieri/Van Hove Phase Shift package \cite{l12}. 

For describing inelastic scatterings, we apply the Lindhard dielectric function \cite{l17prim} together with optical-data models \cite{l19,l21,l14,l15}. The response of the medium to a passing electron, giving an energy loss of $\hbar\om$ and a momentum change $\hbar{\bf q}$, is described in those models
by a complex dielectric function \cite{l17prim} $\eps({\bf q},\om)$. The probability of an energy loss $\hbar\om$ per unit distance travelled by a non-relativistic electron of energy $E$, i.e. the differential inverse mean free path, $\tau(E,\om)$ \cite{l17prim,l17quar,l21,l21prim}, then reads~: 
\eq 
\tau(E,\om)=\frac{1}{\pi E a_0}\,\int_{q_-}^{q_+}\, 
\frac{dq}{q}\,Im[-\eps(q,\om)^{-1}], 
\label{tau} 
\eqx 
where $a_0$ is the Bohr radius, and
\eq 
\hbar q_{\pm}=\sqrt{2m_e}\,
\left(\sqrt{E}\,\pm \,\sqrt{E-\hbar\om}\right) 
\label{qpm} 
\eqx 
are the maximal and minimal values of the allowed momentum transfer to the solid, and $m_e$ denotes the mass of a free electron. The expression for $\hbar q_{\pm}$ assumes that the energy and momentum transfer for an electron moving in the medium is the same as for a free particle in vacuum, i.e. there is no effective mass assumed and no energy gap. Integration of the differential inverse mean free path over the allowed values of $\om$ yields the total inverse inelastic mean free path \cite{ziaja}. It follows from (\ref{tau}) that the only quantity needed to evaluate $\tau(E,\om)$ and $\lambda_{in}(E)$ is the dielectric response function $\eps(q,\om)$. However, most existing data on dielectric response functions were obtained from photon scattering on solids, for which the momentum transfer is zero. The problem is how to predict $\eps(q,\om)$ with $q>0$, knowing only its optical limit ($q=0$) \cite{l21,l21prim}. For that purpose a phenomenological optical model was introduced, where $Im[-\eps(q,\om)^{-1}]$ is expressed via the convolution of $Im[-\eps(q=0,\om)^{-1}]$ with some profile function of $q$ and $\om$. We used two such models in the following. The optical model by Ashley \cite{l21,l21prim} includes exchange between the incident electron and the electron in the medium, which is modeled in analogy with the structure of the non-relativistic M\o{}ller cross-section. The Tanuma, Powell and Penn model (TPP-2) \cite{l14} was adopted for calculating the differential inelastic mean free path and inelastic mean free path of electrons in the solid. The energy loss function for diamond used in these calculations was taken from Ref.\ \cite{ziaja}. Diamond shows a dominant peak in its electron-loss function, corresponding to well-defined volume plasmons \cite{l19}. This means that the Lindhard dielectric function describes this solid satisfactorily \cite{l14}, at least in the resonance region that is far from the minimum energy needed to make an electron-hole pair.

For the sake of simplicity, we only consider the system at low dose rates, and very early in an exposure when the medium is generally still neutral. Thus we neglect long-range Coulomb interactions, in particular electron-electron interactions, and therefore assume that electron trajectories follow straight lines between consecutive scatterings. Consequently, we do not simulate the motion of the holes and neglect any electron-hole interaction.
Effects of hole propagation, i.\ e.\ possible inelastic scatterings of the holes and subsequent ionization, are neglected as well, since these 
processes do not contribute significantly to the total ionization rate. 
Due to the low impact energy of primary electrons ($E\sim 250$ eV), we neglect interactions of the electrons with inner shells of the atoms (cf.\ \cite{l29,l30,models1,models2,models3,models4,akker}). The interactions with valence electrons are modelled as interactions with the band in the solid. For electrons in the conduction band of kinetic energy lower than $E_{elast}\sim14$ eV in Ashley's, and $E_{elast}\sim9$ eV in the TPP-2 model, the inelastic mean free path predicted by these models was very large. Therefore we assumed that below these energies electrons scatter elastically only. Interactions of conduction band electrons with lattice phonons, which are important for low kinetic energies ($E<10$ eV) were neglected \cite{models1}.  Following Ashley's model and the TPP-2 model, we did not allow any form of electronic recombination, so electrons excited into the conduction band stayed there
during the simulations.
%
%
\begin{figure}[t] 
\centerline{\epsfig{figure=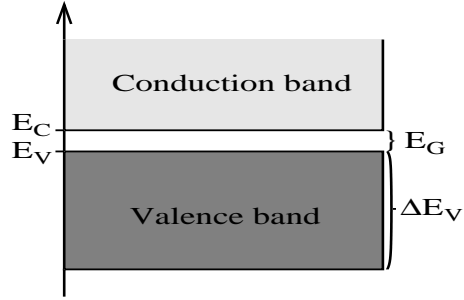,width=6cm, height=4cm}} 
\caption{{\footnotesize Band structure of diamond at $T=300$ K. The zero of the energy scale lies in the middle of the band gap.
The position of the bottom of the conduction band is $E_C=E_G/2$, and the 
band gap equals $E_G=5.46$ eV. The energy $E_V$, $E_V=-E_G/2$, is the energy 
at the top of the valence band. The width of the valence band is estimated to be $\Delta E_V\sim 23$ eV. Numerical values were taken from Ref.\ \cite{diamond1}.}}
\label{f1} 
\end{figure} 
\noindent
\begin{figure}[t] 
\epsfig{figure=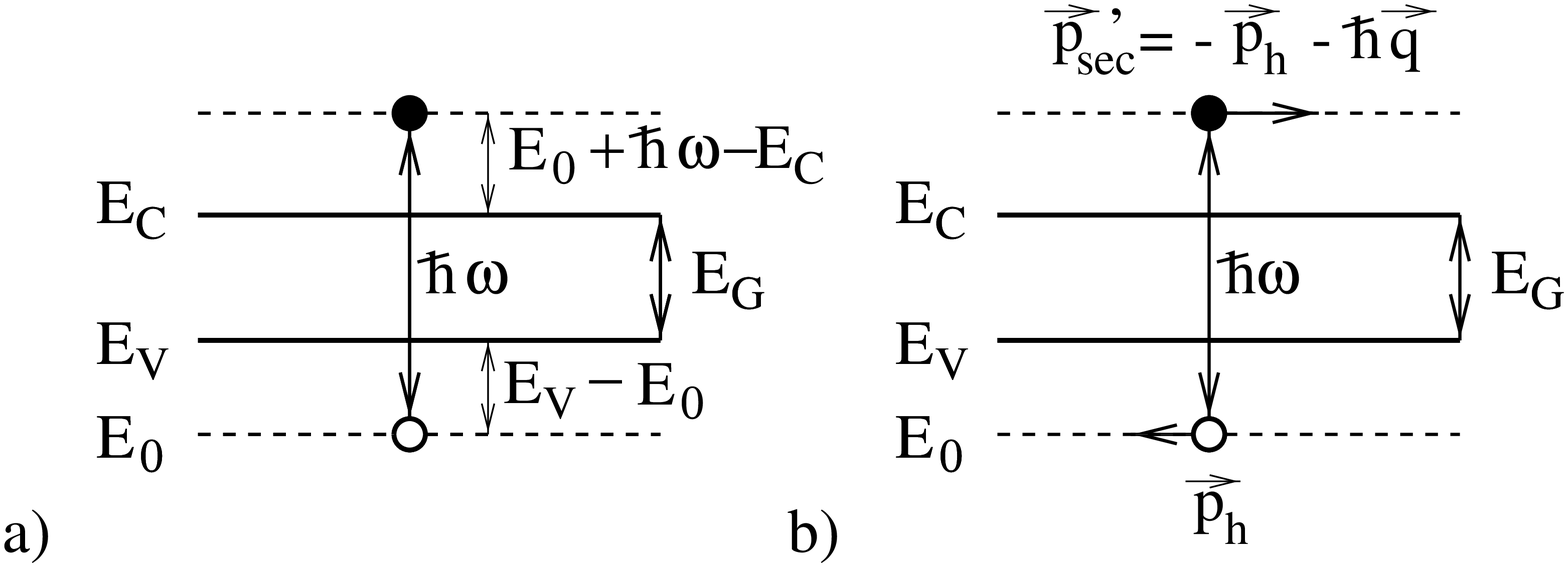,width=14cm}\hfill\mbox{}\\
\caption{{\footnotesize Excitation of an electron-hole pair in an insulator/semiconductor as modelled: {\bf (a)} the kinetic energy of the 
excited electron is $E_0+\hbar\omega-E_C$, where $E_0$ is the initial energy of the electron in the valence band, and $\hbar\omega$ is the energy transfer; 
{\bf (b)} Momentum of the electron equals 
${\bf p}_{sec}^{\prime}=-{\bf p}_{h}-\hbar{\bf q}$, where ${\bf p}_{h}$ is the momentum of the hole remaining in the valence band, and $\hbar {\bf q}$ denotes momentum transfer. The kinetic energy of the hole is $E_V-E_0$. The velocity and the momentum of a hole point in opposite directions \cite{l6}.  
}}
\label{f11} 
\end{figure} 
\noindent
\begin{figure}[t]
a)\epsfig{width=7cm, file=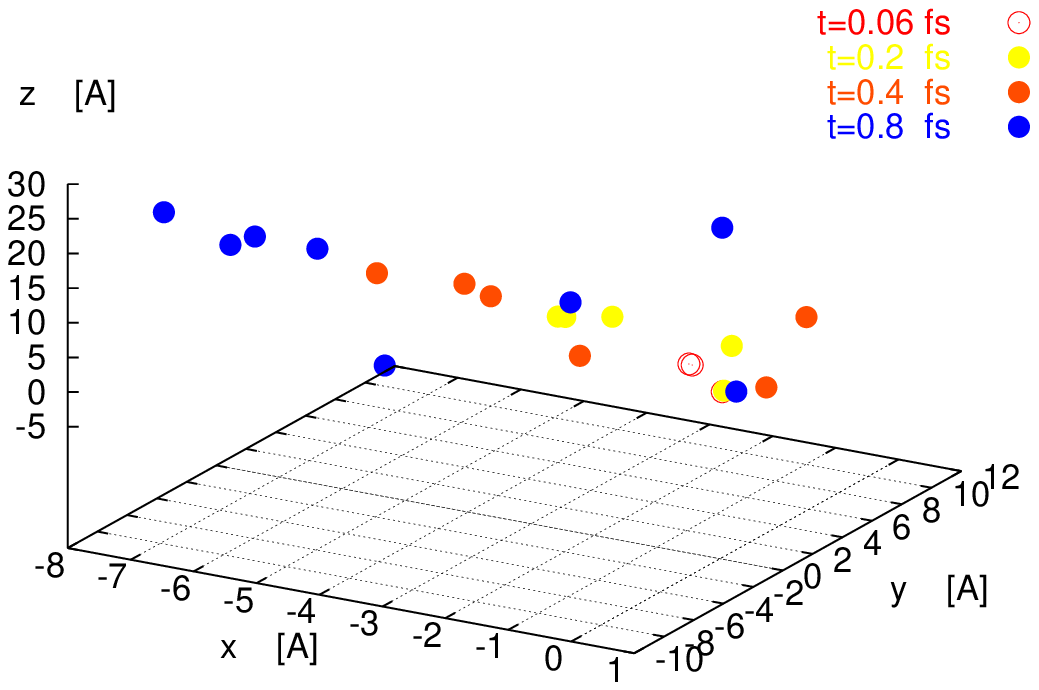}\hfill\mbox{}
b)\epsfig{width=7cm, file=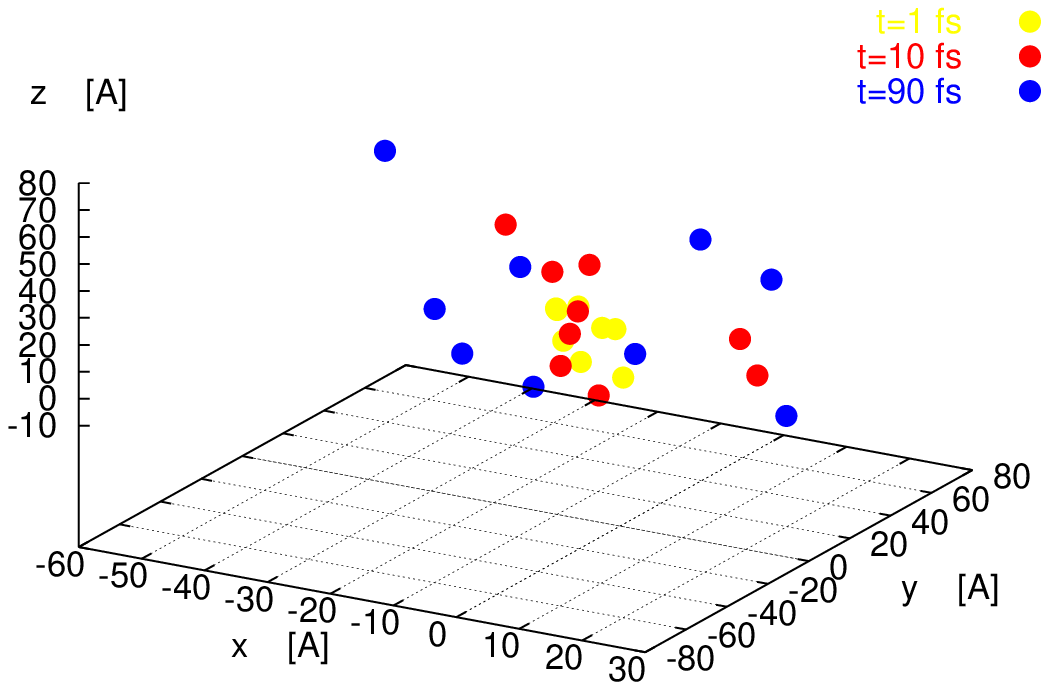}\hfill\mbox{}\\
\caption{{\footnotesize Evolution of secondary electron cascade in diamond at different time steps in the TPP-2 model. The coordinates of impact electron at $t=0$ fs are ${\bf x}=(0,0,0)$ and ${\bf v}=(0,v_y,0)$, where $v_y=\sqrt{2E/m_e}$. Symbols denote electrons at different time points $t=0.06, 0.2, 0.4, 0.8, 1, 10, 90$ fs respectively.}}
\label{f2}
\end{figure}
\noindent
\begin{figure}[t]
\begin{center}
a)\epsfig{width=9cm, file=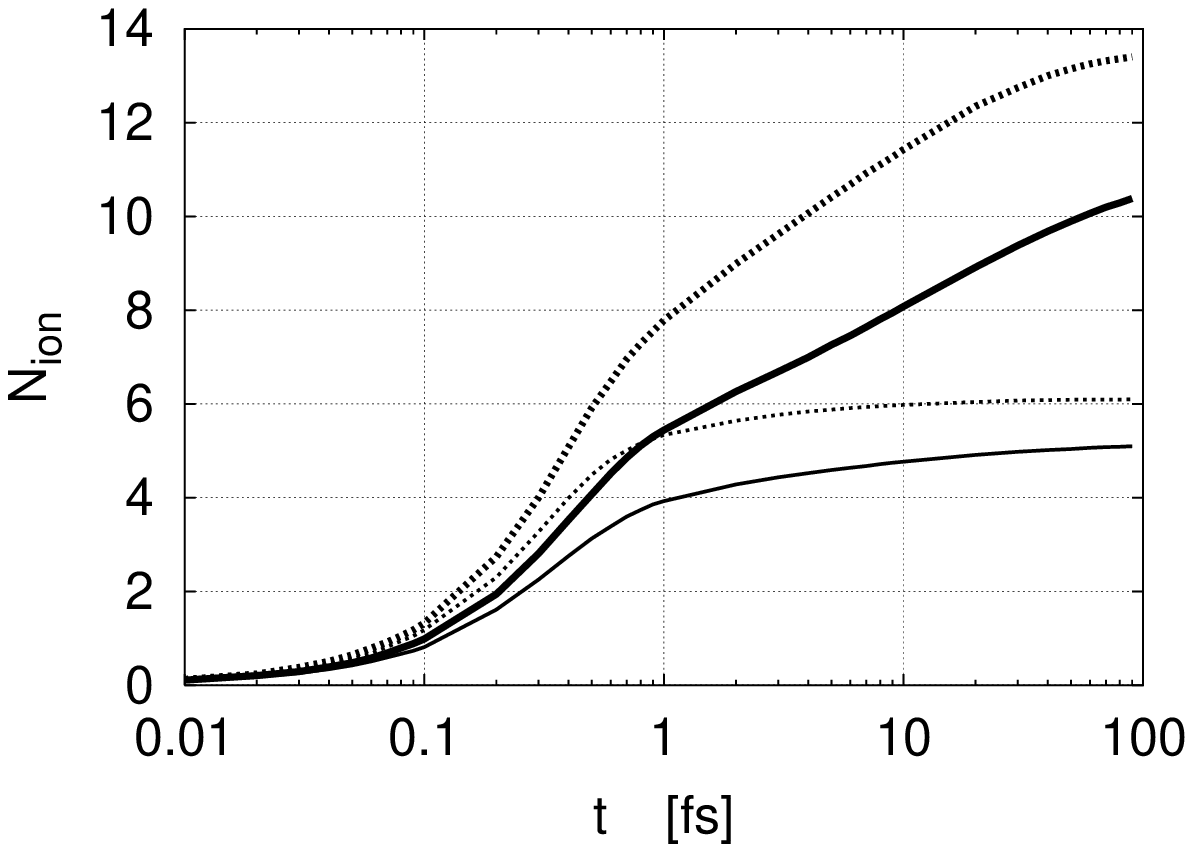}\\
b)\epsfig{width=9cm, file=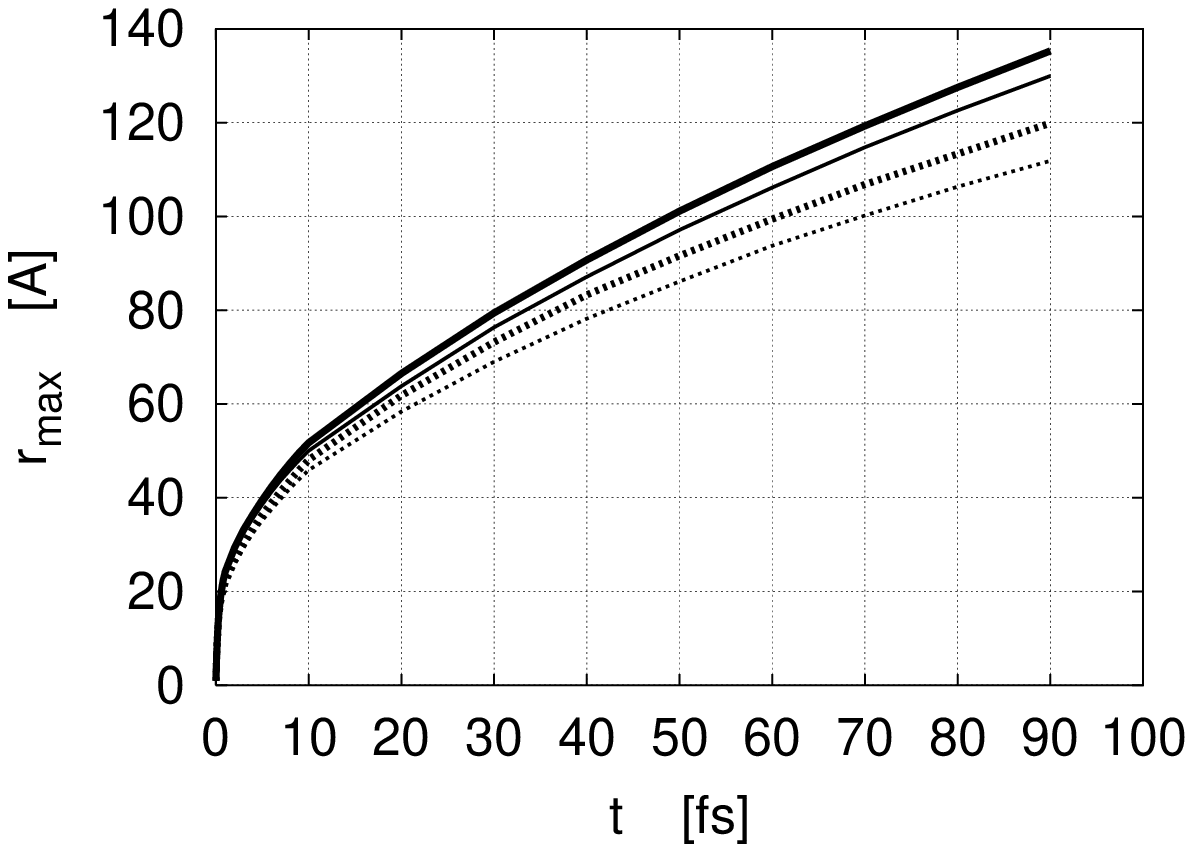}\\
c)\epsfig{width=9cm, file=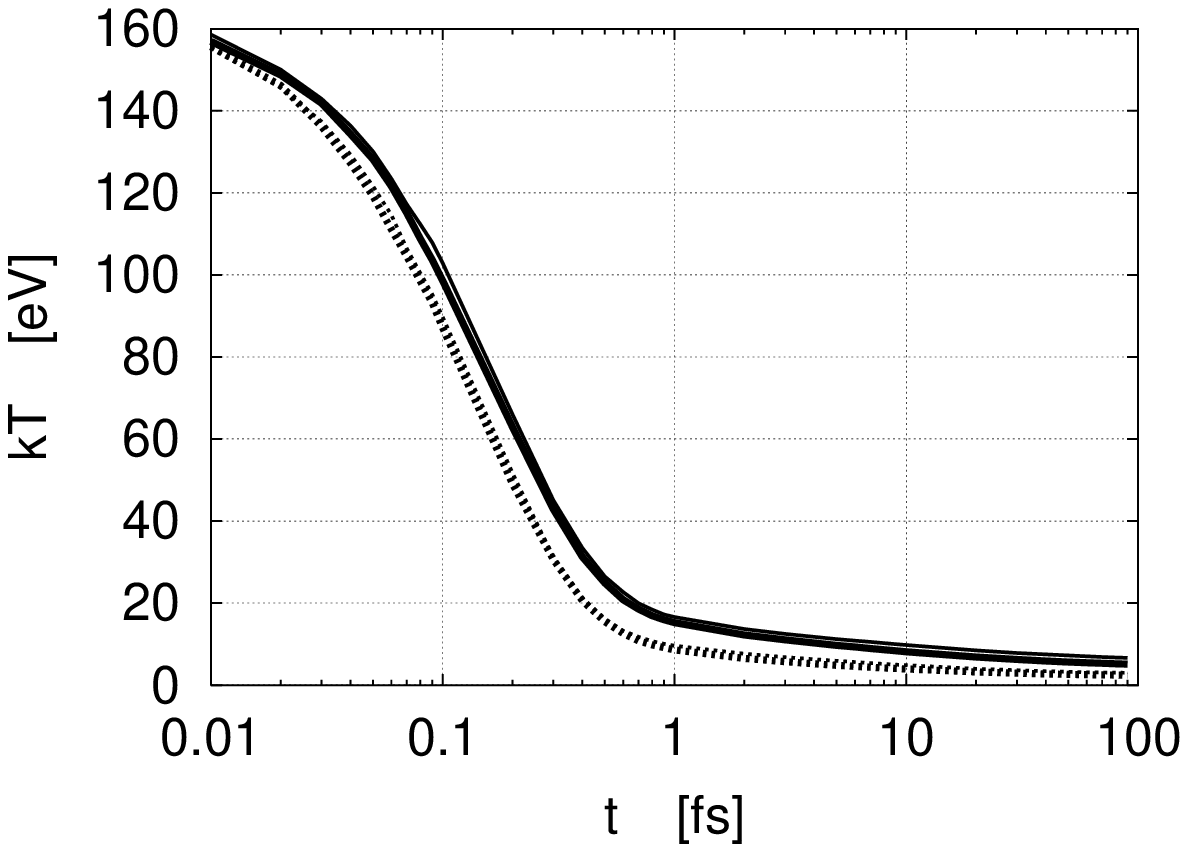}\\
\end{center}
\caption{{\footnotesize {\bf (a)} Average ionization rate ($N_{ion}$ vs. time); {\bf (b)} Average range of cascade electrons $r_{max}$ (maximal radius) vs. time, $t$. Position of primary electron at $t=0$ fs is at $r=0$ \AA ; {\bf (c)} Average temperature $kT$ of electron gas vs. time. Solid lines correspond to the results obtained from Ashley's model, dashed lines show results with
the TPP-2 model. {\bf Thick} curves show results with {\bf no} energy transfer allowed to the lattice. {\bf Thin} curves show the results with energy transfer to the lattice.
}}
\label{f3}
\end{figure}
\noindent
\begin{figure}[t]
\begin{center}
a)\epsfig{width=9cm, file=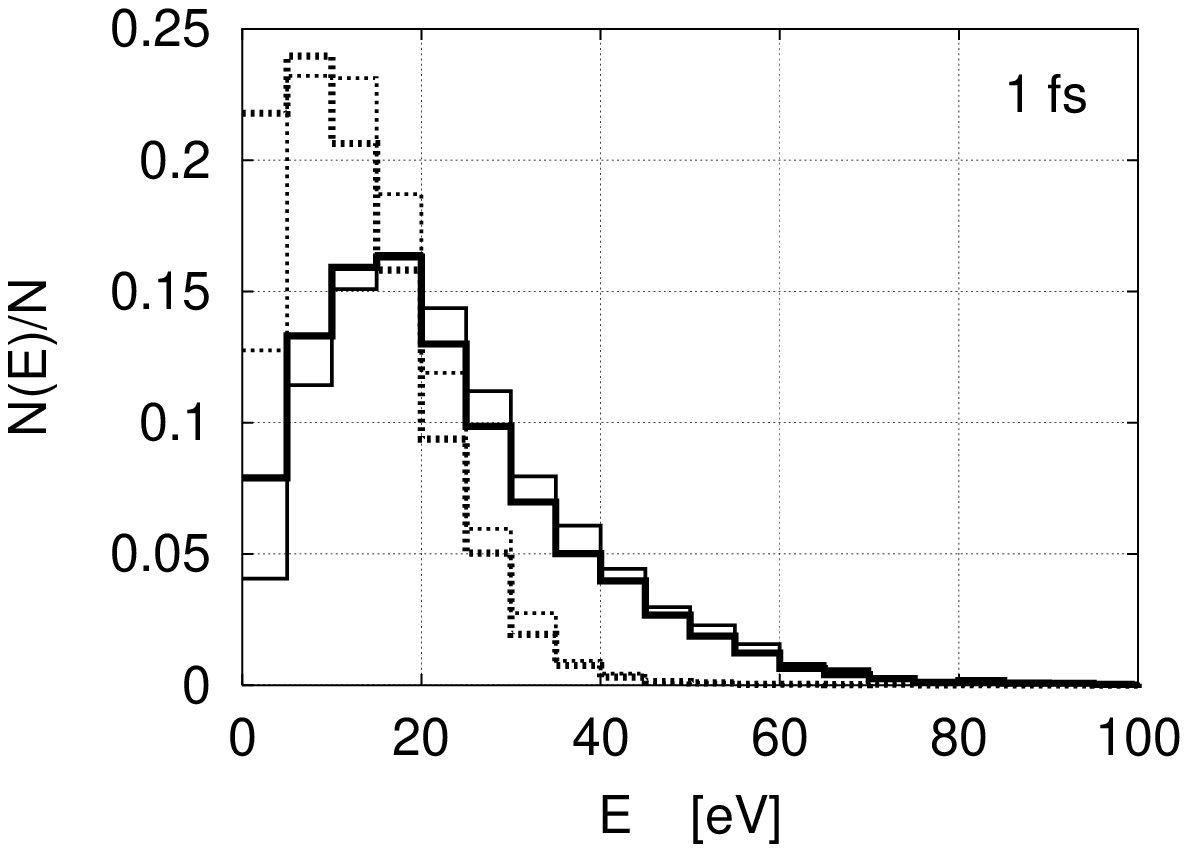}\\
b)\epsfig{width=9cm, file=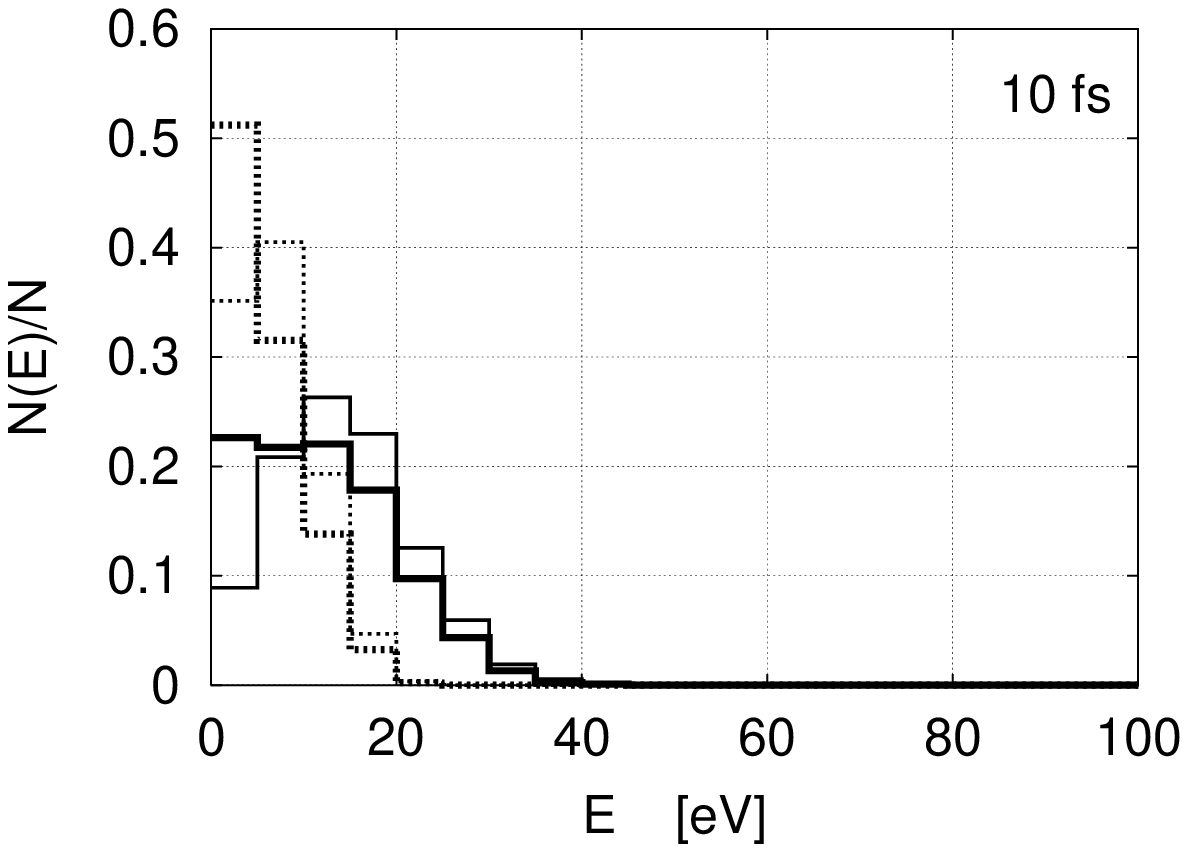}\\
c)\epsfig{width=9cm, file=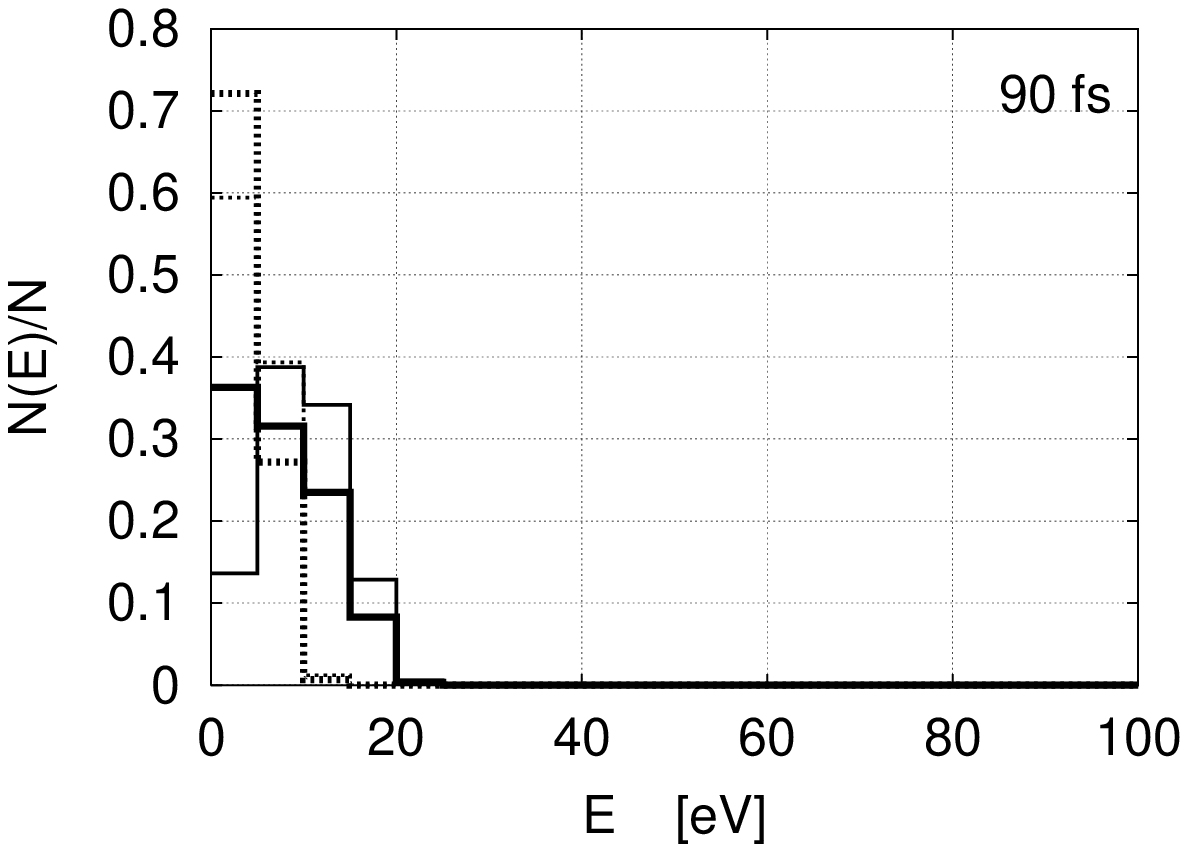}\\
\end{center}
\caption{{\footnotesize Energy distribution $N(E)/N$ among electrons (histogram) at {\bf (a)} $t=1$ fs; {\bf (b)} $t=10$ fs; and {\bf (c)} $t=90$ fs. Solid lines correspond to results obtained from Ashley's model, dashed lines show  results from the TPP-2 model. {\bf Thick} lines show the results when {\bf no} energy transfer to the lattice is allowed. {\bf Thin} lines correspond to the results with energy transfer to the lattice. 
}}
\label{f4}
\end{figure}
%
\noindent
\begin{figure}[t]
\epsfig{width=15cm, file=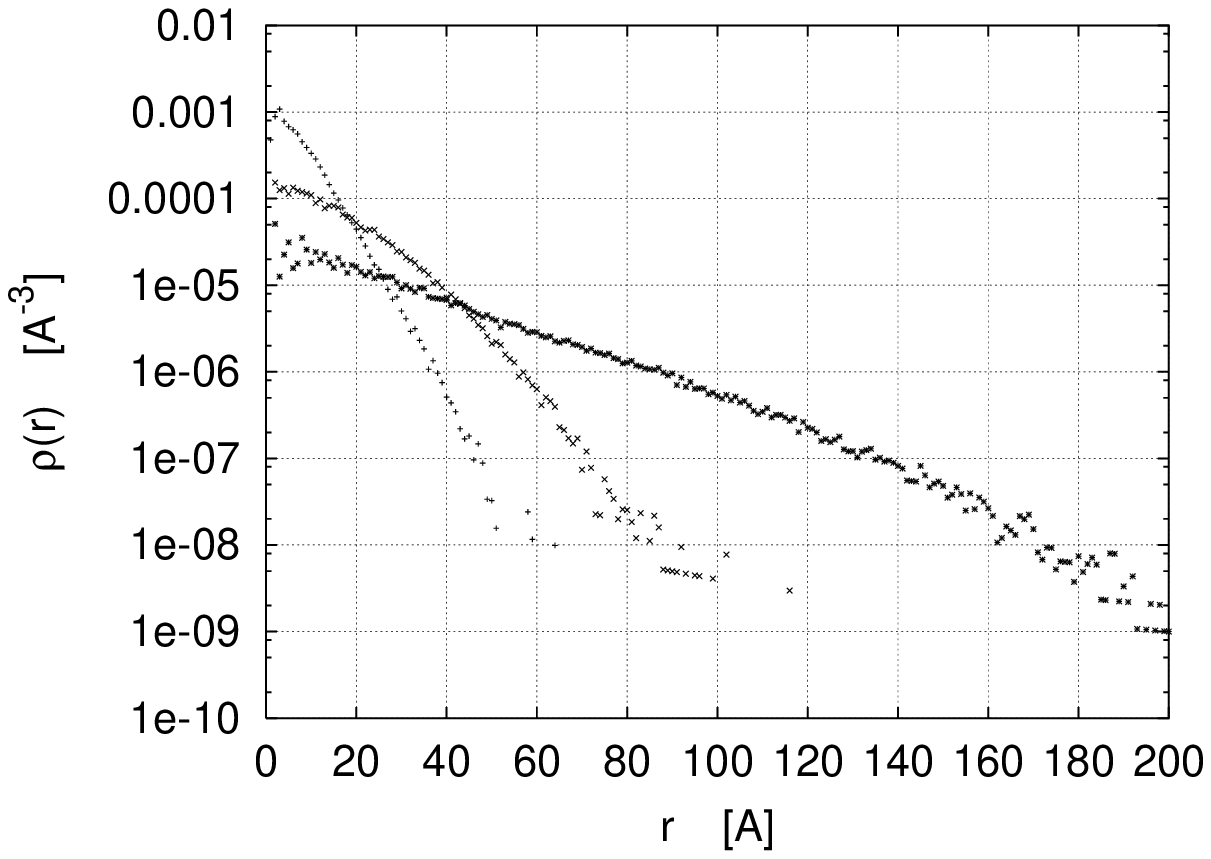}\hfill\mbox{}\\
\epsfig{width=15cm, file=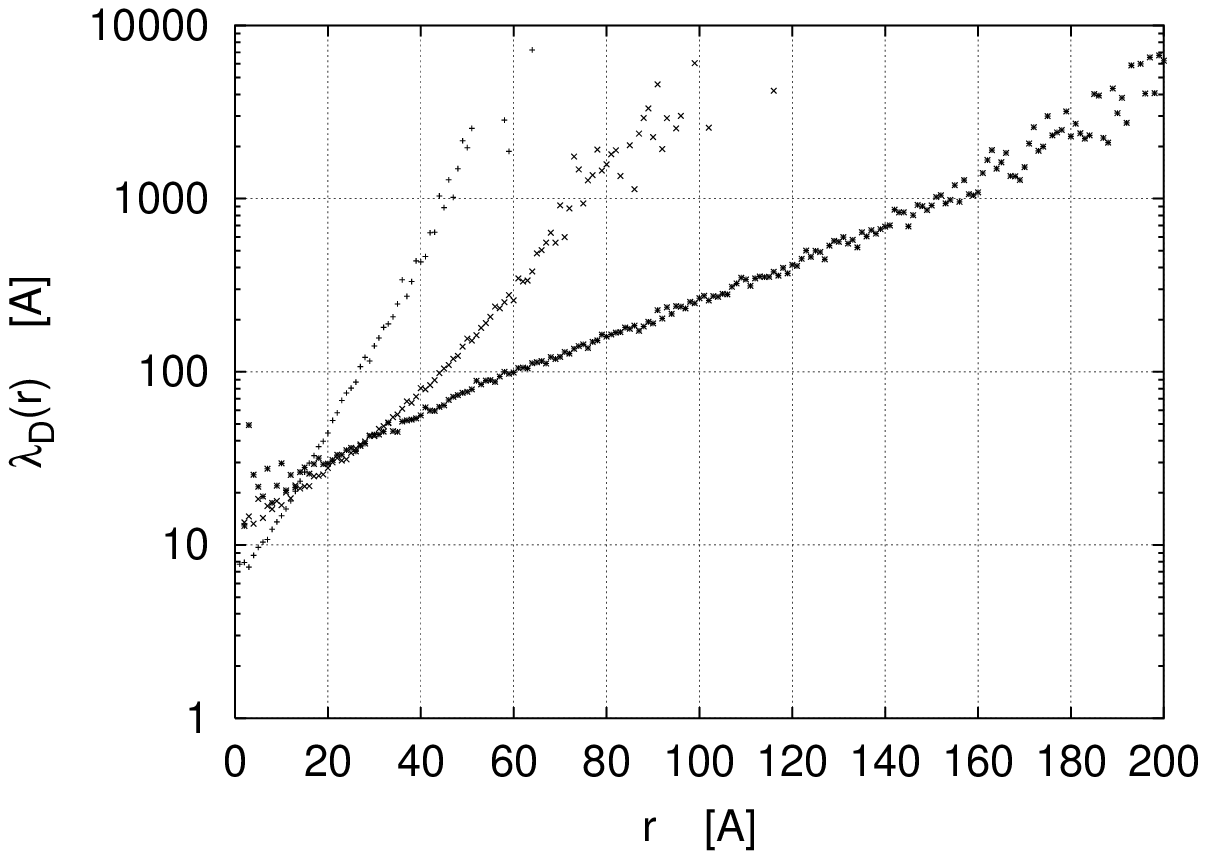}\hfill\mbox{}\\
\caption{Electron density, $\rho$ vs. radius, $r$, (upper plot) and
Debye length, $\lambda_D$, vs. radius, $r$, (lower plot) at times: $t=1$ fs 
(stars), $t=10$ fs ($\times$), $t=90$ fs (crosses). Results with 
Ashley's model with no energy loss to the lattice.}
\label{fig1}
\end{figure}
\noindent
\begin{figure}[t]
\epsfig{width=15cm, file=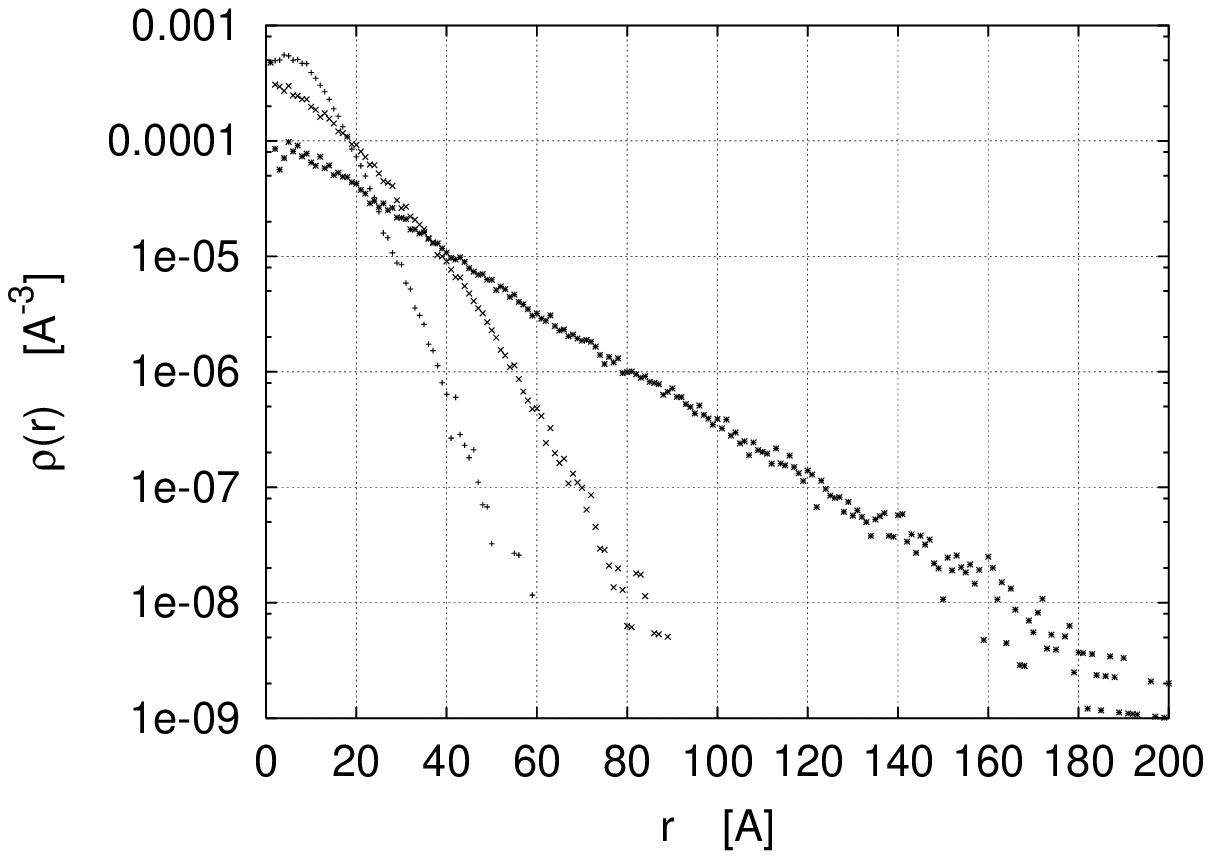}\hfill\mbox{}\\
\epsfig{width=15cm, file=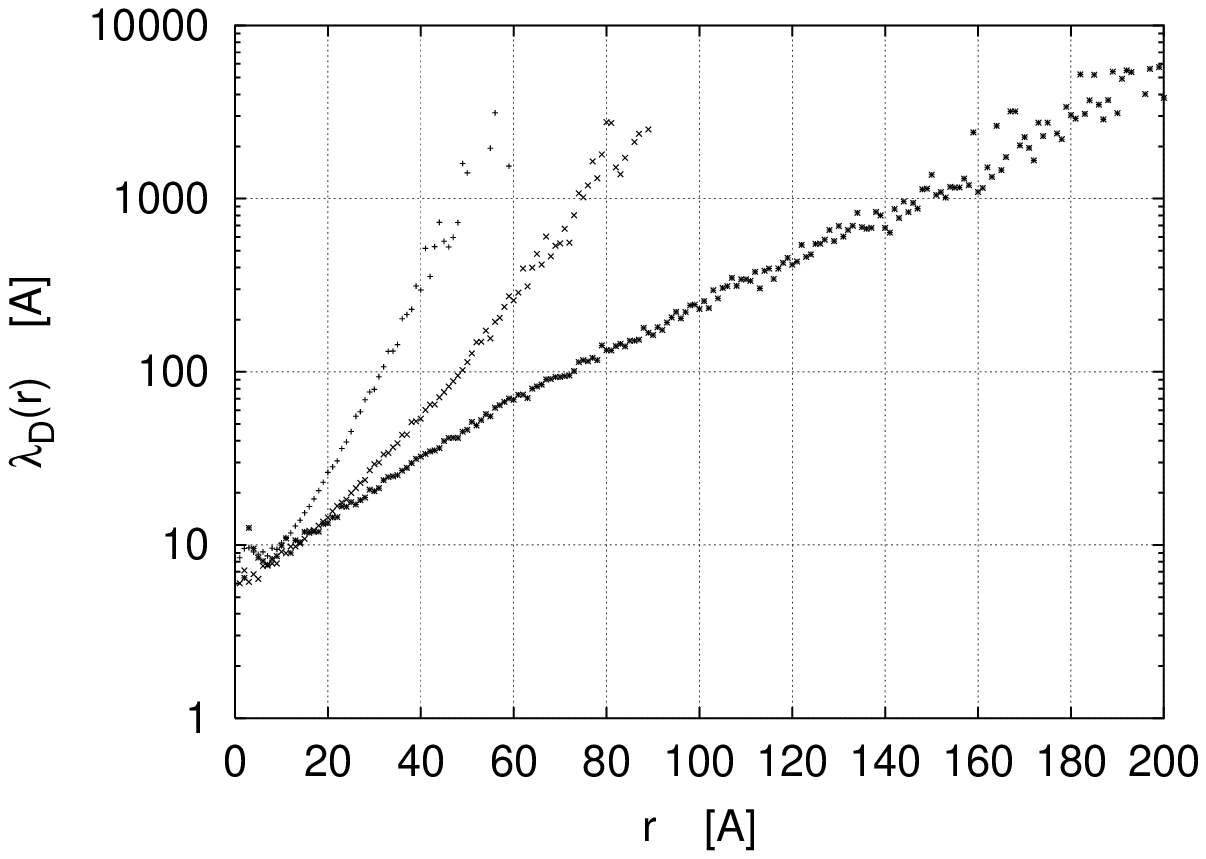}\hfill\mbox{}\\
\caption{Electron density, $\rho$ vs. radius, $r$, (upper plot) and
Debye length, $\lambda_D$, vs. radius, $r$, (lower plot) at times: $t=1$ fs
(stars), $t=10$ fs ($\times$), $t=90$ fs (crosses). Results with 
the TPP-2 model with no energy loss to the lattice.}
\label{fig2}
\end{figure}
\noindent
\begin{figure}[t]
\epsfig{width=15cm, file=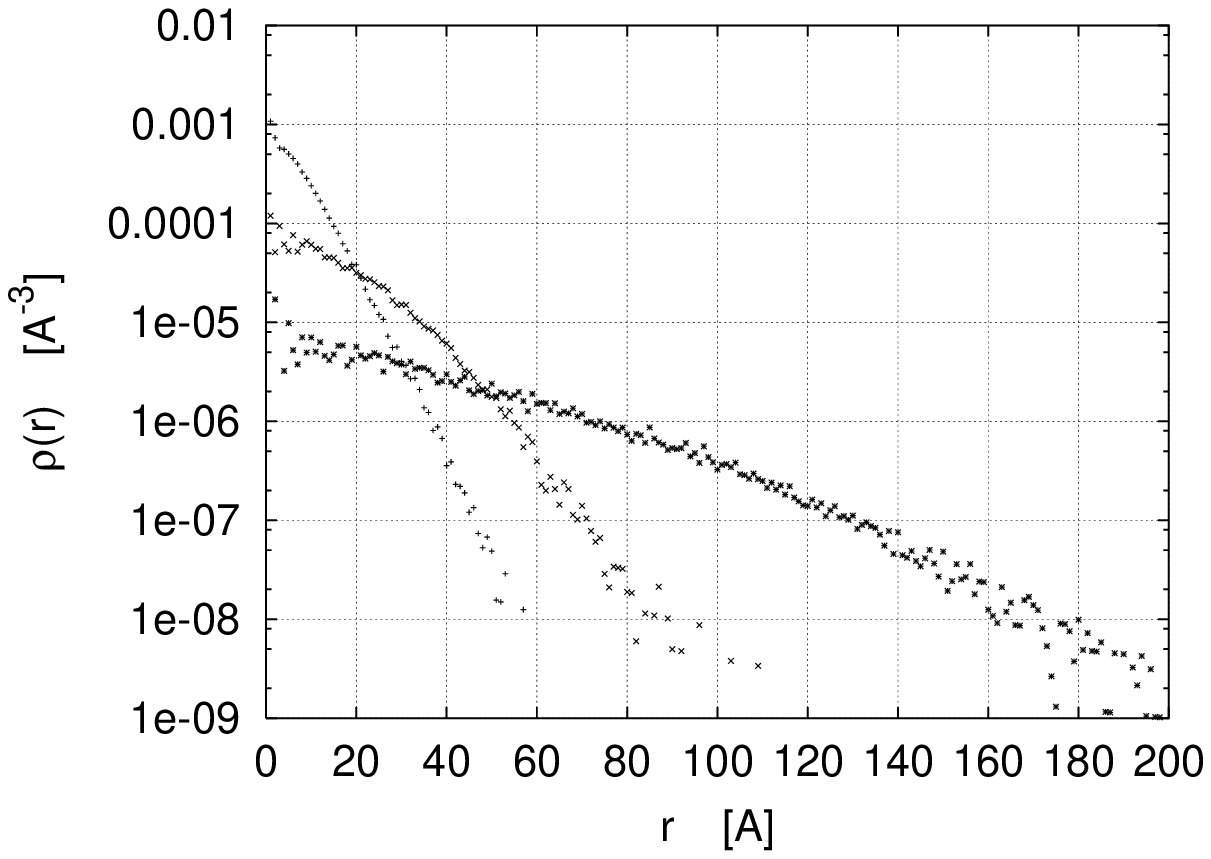}\hfill\mbox{}\\
\epsfig{width=15cm, file=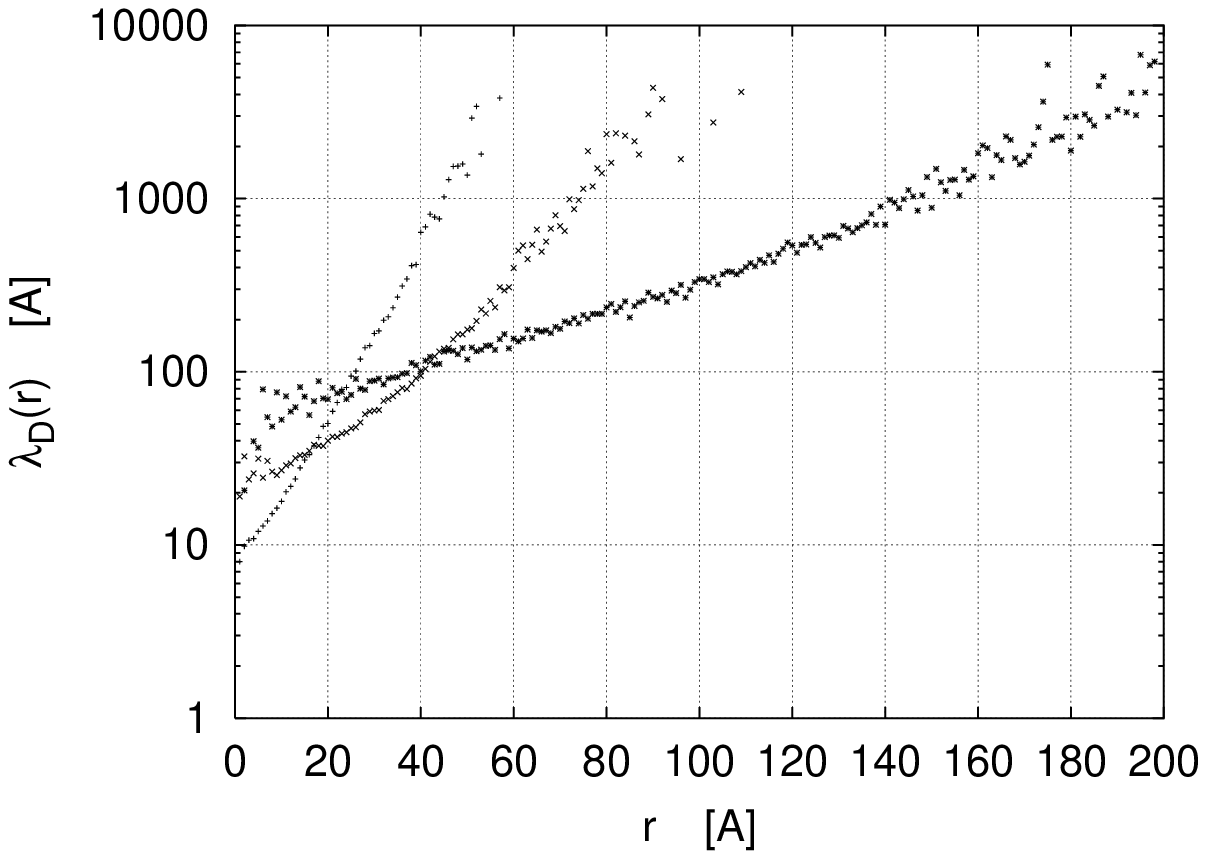}\hfill\mbox{}\\
\caption{Electron density, $\rho$ vs. radius, $r$, (upper plot) and
Debye length, $\lambda_D$, vs. radius, $r$, (lower plot) at times: $t=1$ fs
(stars), $t=10$ fs ($\times$), $t=90$ fs (crosses). Results with 
Ashley's model with energy loss to the lattice allowed.}
\label{fig3}
\end{figure}
\noindent
\begin{figure}[t]
\epsfig{width=15cm, file=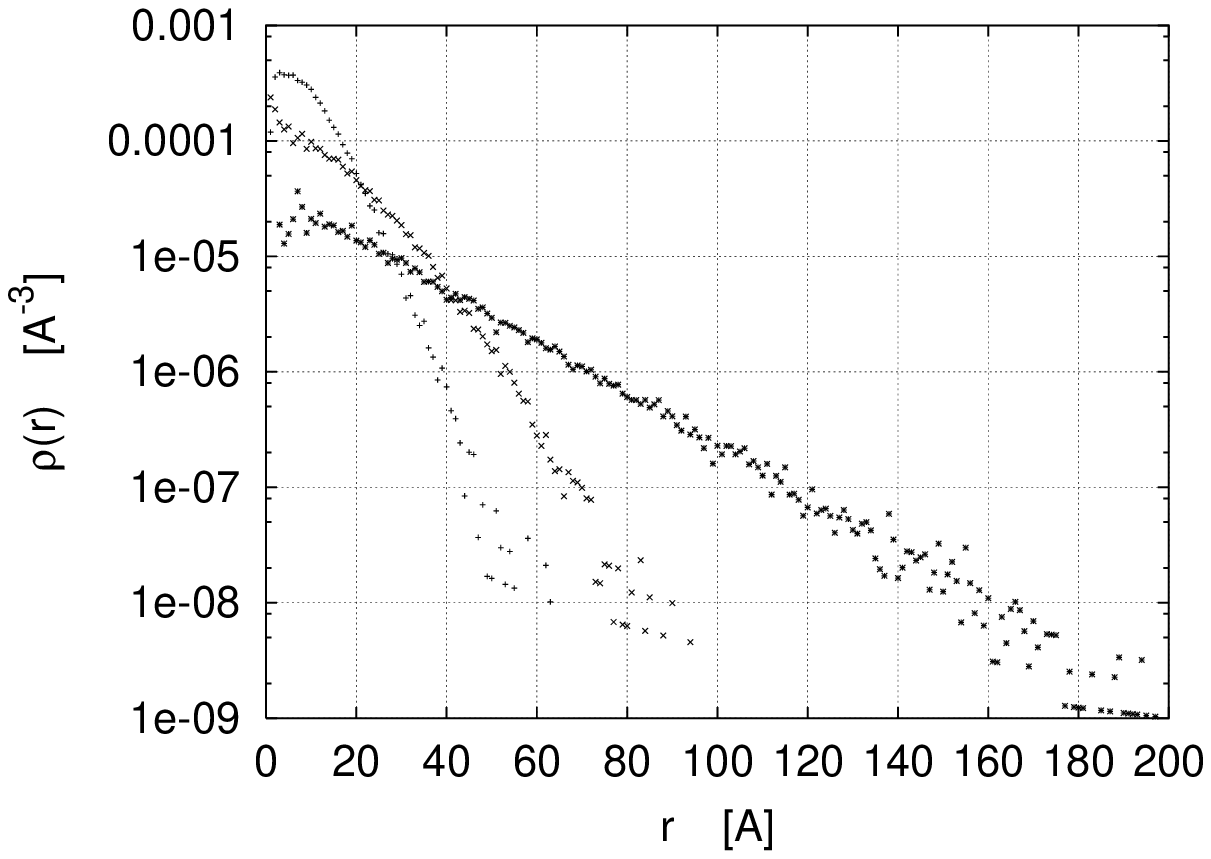}\hfill\mbox{}\\
\epsfig{width=15cm, file=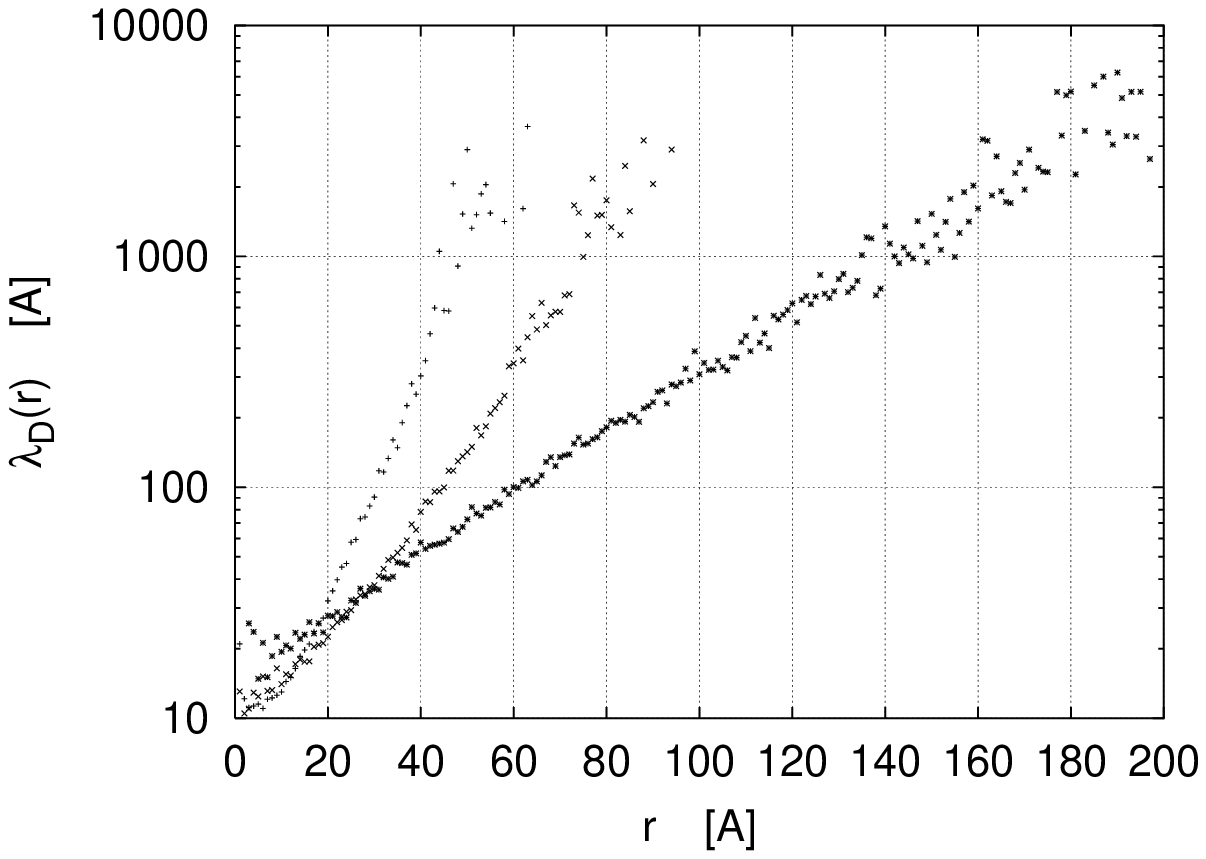}\hfill\mbox{}\\
\caption{Electron density, $\rho$ vs. radius, $r$, (upper plot) and
Debye length, $\lambda_D$, vs. radius, $r$, (lower plot) at times: $t=1$ fs
(stars), $t=10$ fs ($\times$), $t=90$ fs (crosses). Resul(lower plot)ts with
the TPP-2 model with energy loss to the lattice allowed.}
\label{fig4}
\end{figure}
%
%
\noindent
\begin{figure}[t]
\begin{center}
a)\epsfig{width=10cm, file=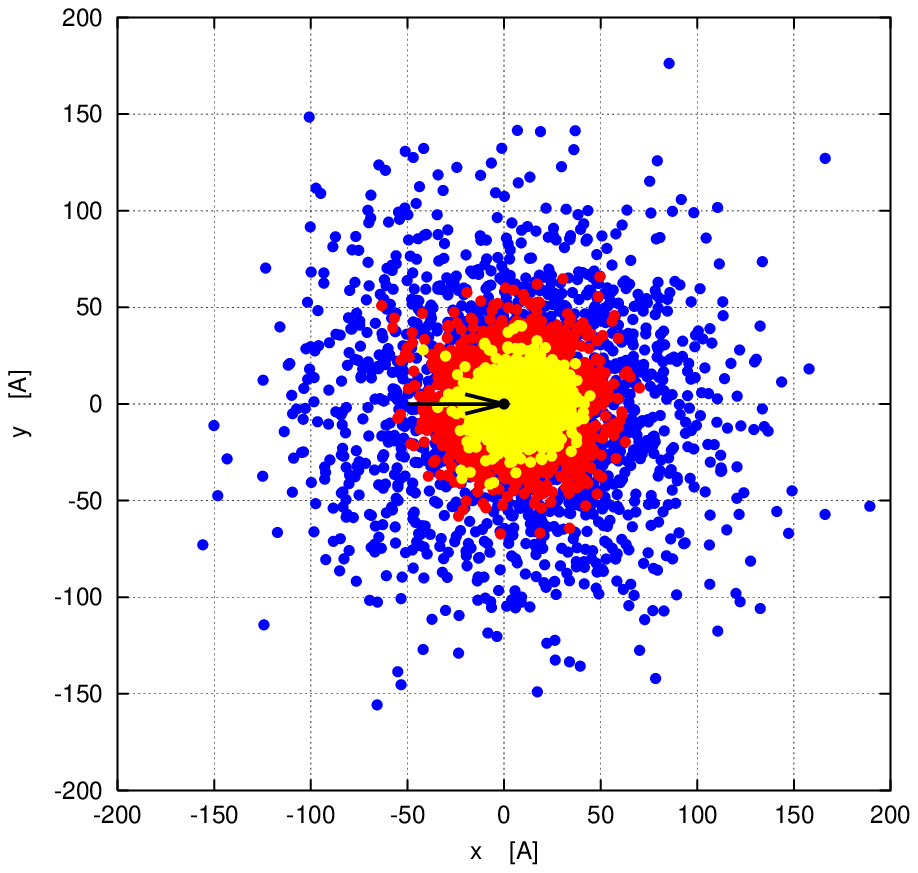}\\
b)\epsfig{width=10cm, file=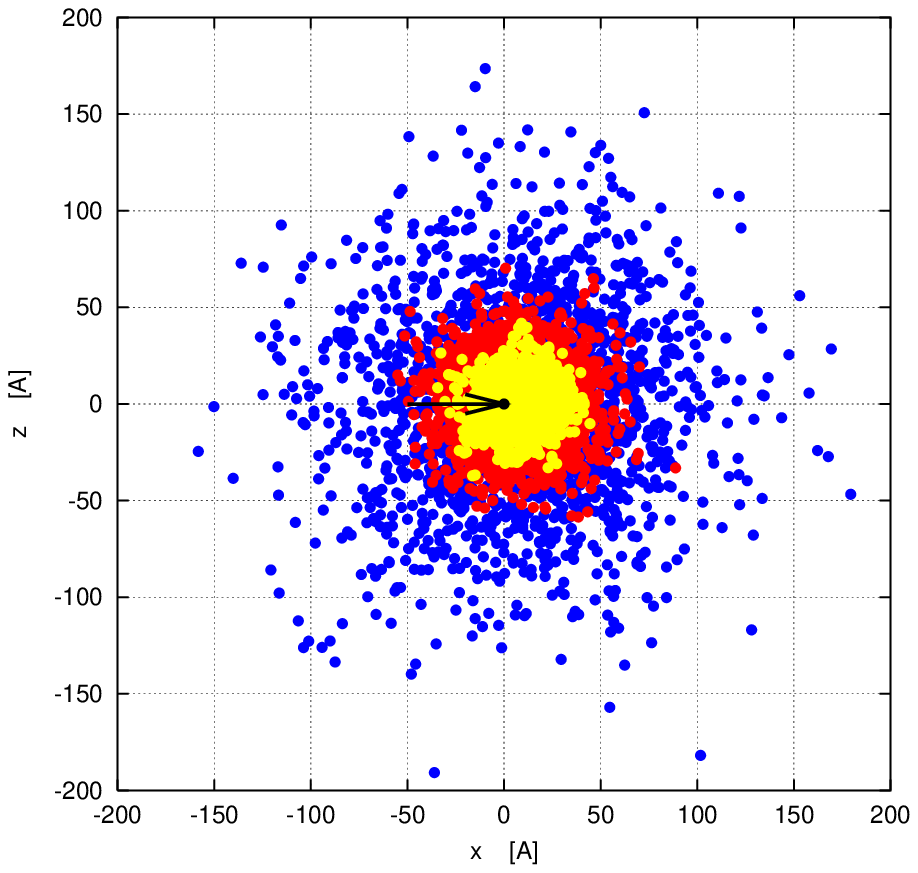}\\
c)\epsfig{width=10cm, file=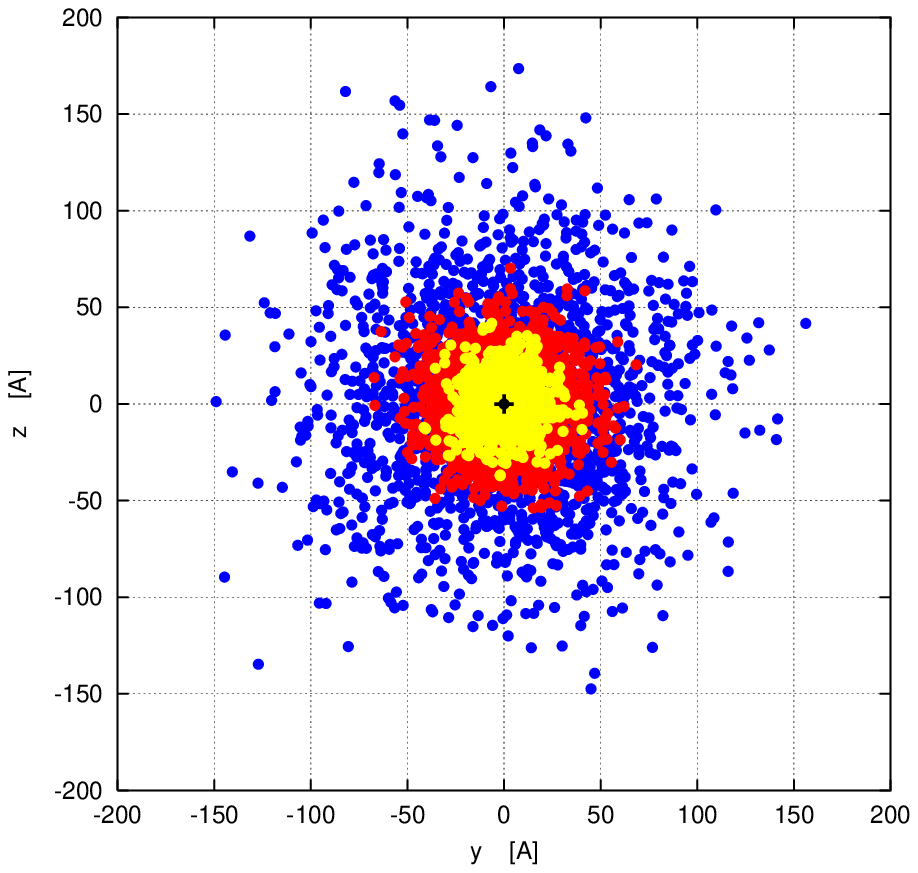}\\
\end{center}
\caption{{\footnotesize Transverse sections of the electron cloud (Ashley's model with the energy transfer to the lattice allowed) in planes: {\bf (a)} XY; {\bf (b)} XZ; {\bf (c)} YZ. The thickness of the section is $40$ \AA . Velocity of the impact electron at $t=0$ fs points towards the X-axis. 
}}
\label{f5}
\end{figure}
\newpage
\noindent
\begin{figure}[t]
\begin{center}
a)\epsfig{width=10cm, file=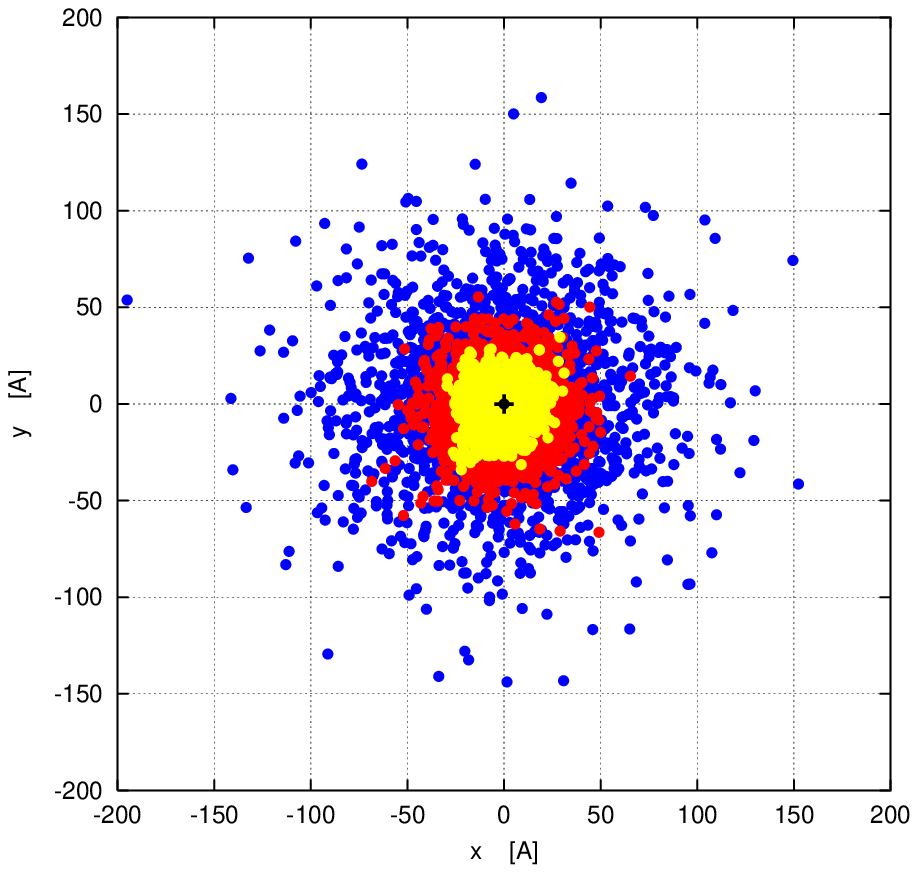}\\
b)\epsfig{width=10cm, file=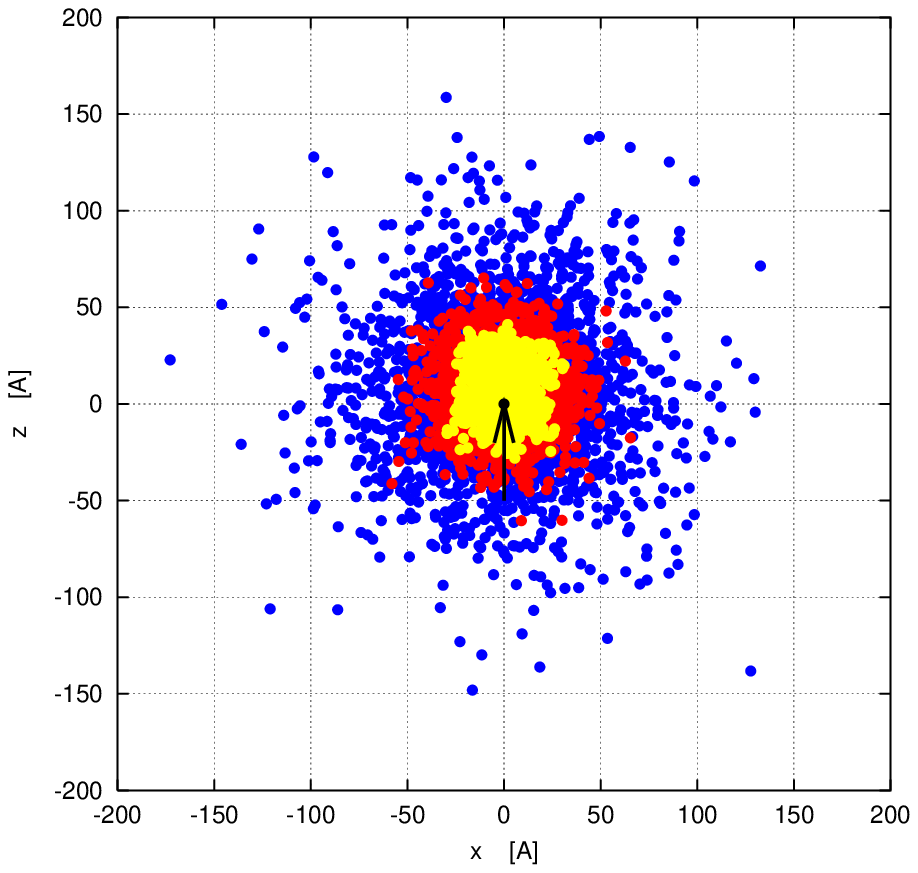}\\
c)\epsfig{width=10cm, file=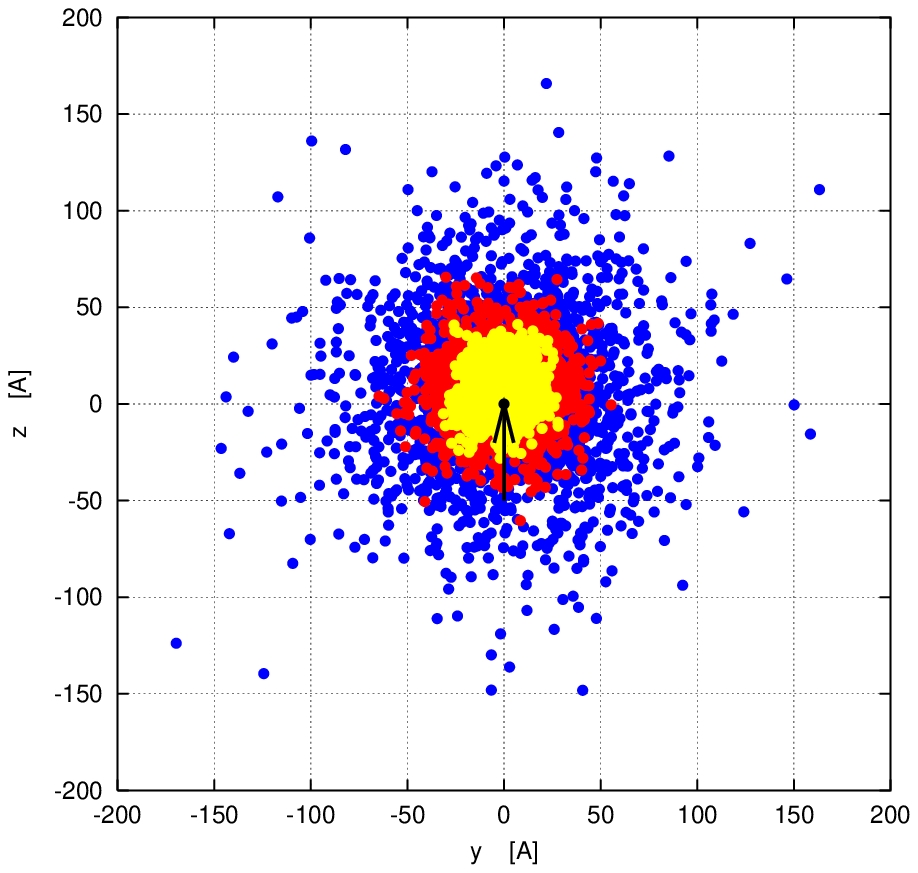}\\
\end{center}
\caption{{\footnotesize Transverse sections of the electron cloud (the TPP-2 model with the energy transfer to the lattice allowed) in planes: {\bf (a)} XY; {\bf (b)} XZ; {\bf (c)} YZ. The thickness of the section is $40$ \AA . Velocity of the impact electron at $t=0$ fs points towards the Z-axis. 
}}
\label{f6}
\end{figure}
\noindent
\begin{figure}[t]
\begin{center}
a)\epsfig{width=10cm, file=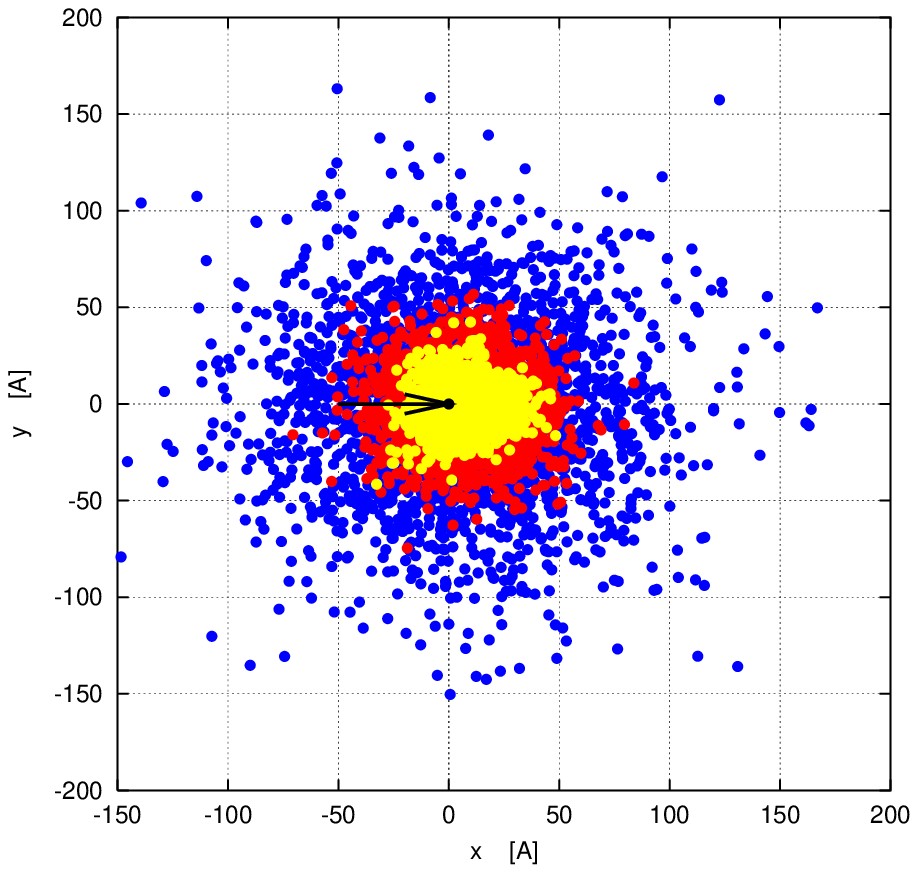}\\
b)\epsfig{width=10cm, file=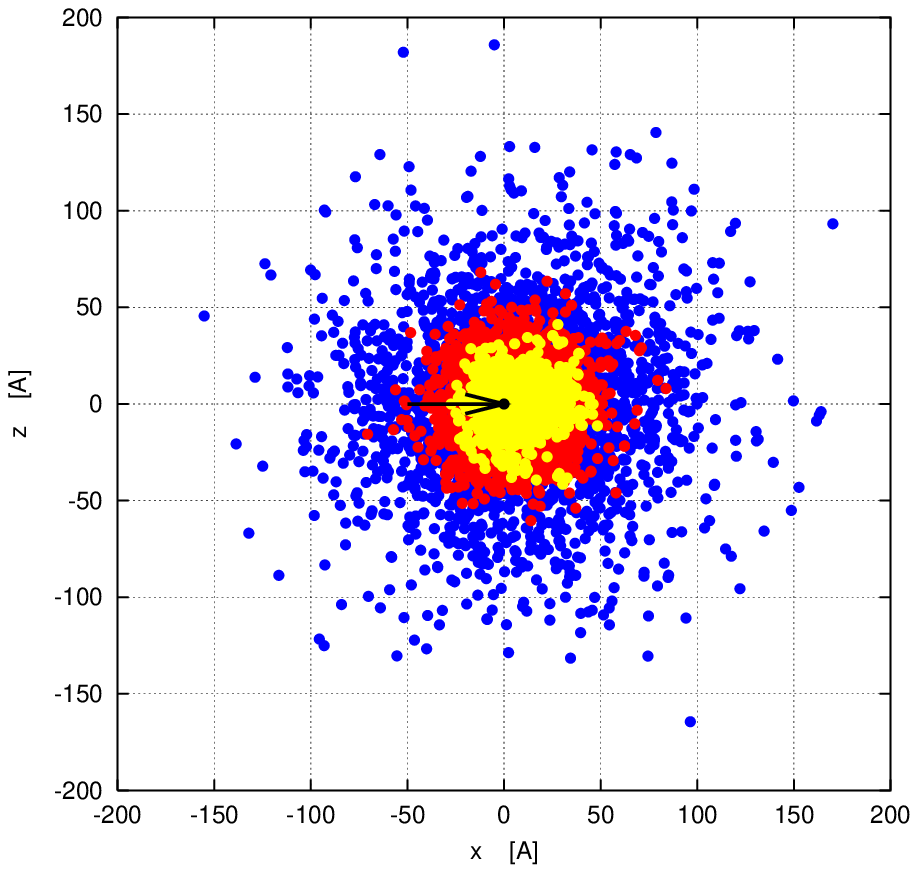}\\
c)\epsfig{width=10cm, file=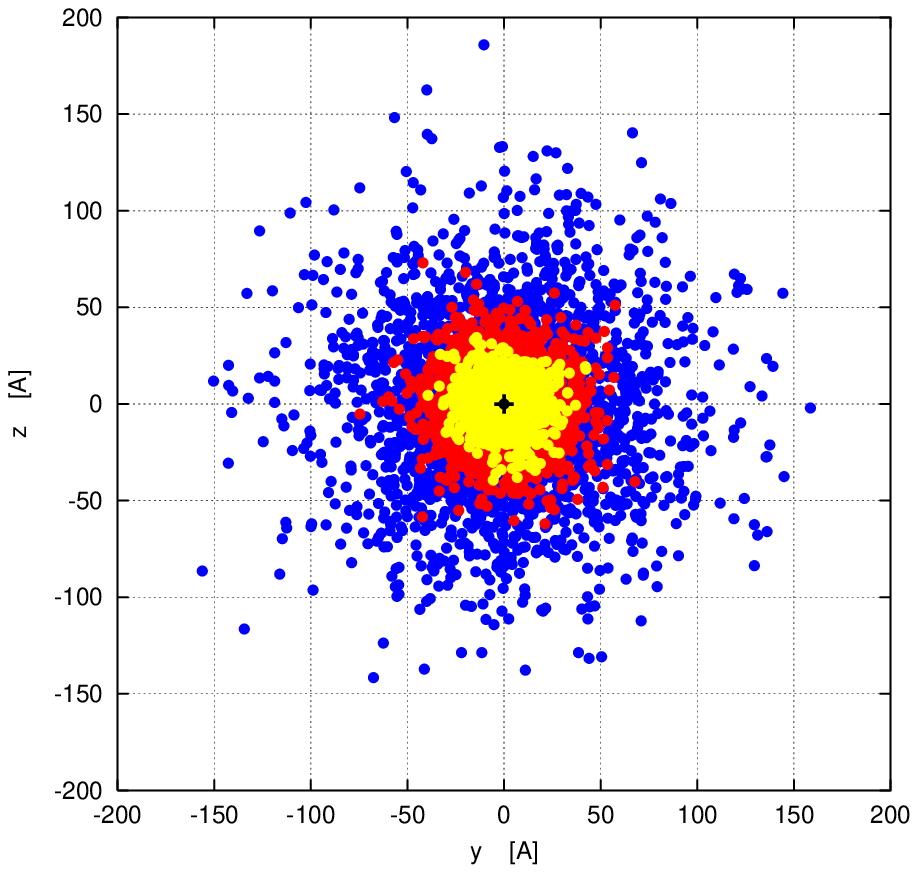}\\
\end{center}
\caption{{\footnotesize Transverse sections of the electron cloud obtained from Ashley's model when no energy transfer to the lattice is allowed in planes: {\bf (a)} XY; {\bf (b)} XZ; {\bf (c)} YZ. The thickness of the section is $40$ \AA . Velocity of the impact electron at $t=0$ fs points towards the X-axis. 
}}
\label{f7}
\end{figure}
\noindent
\begin{figure}[t]
\begin{center}
a)\epsfig{width=10cm, file=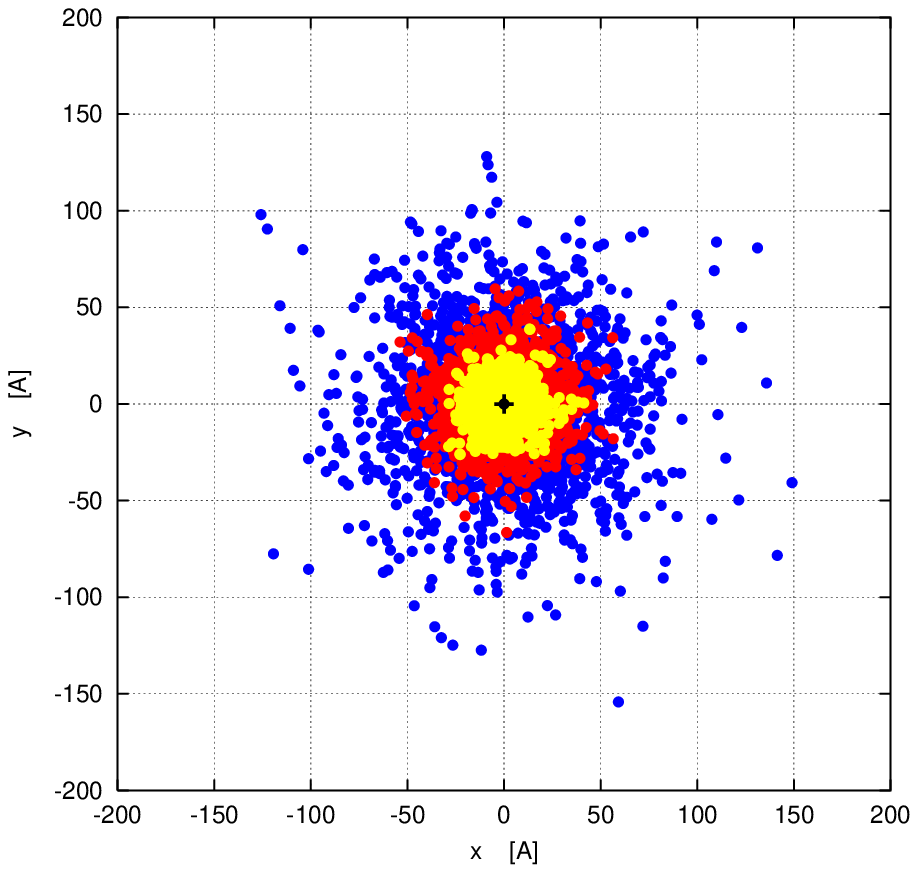}\\
b)\epsfig{width=10cm, file=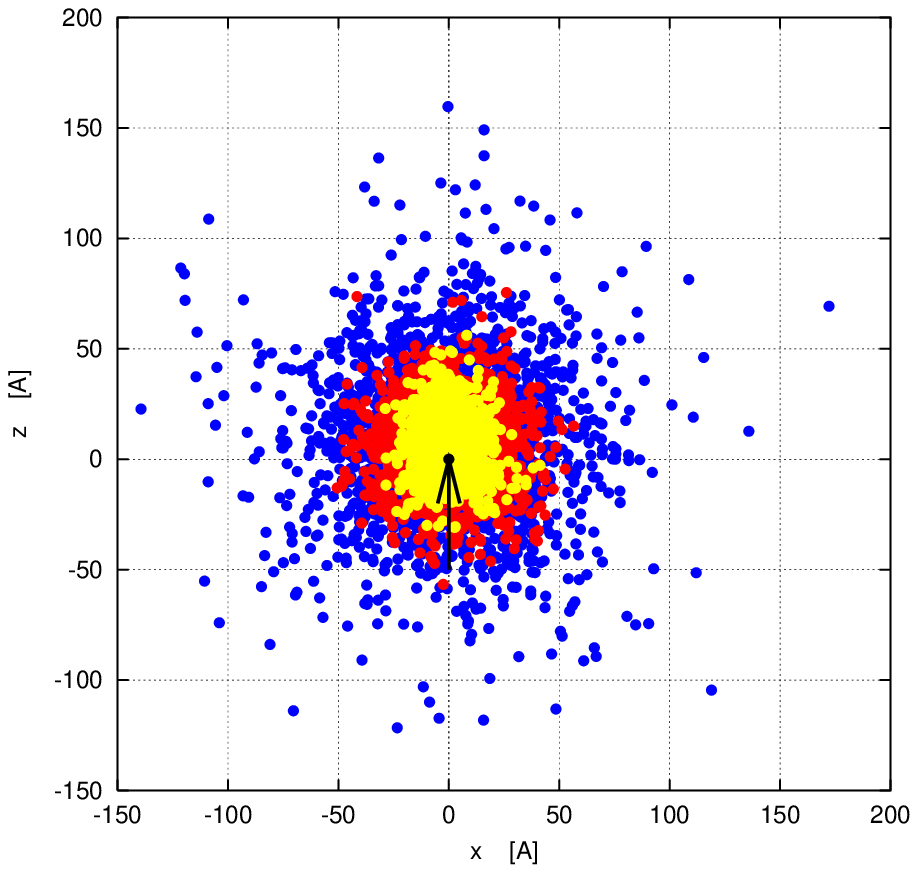}\\
c)\epsfig{width=10cm, file=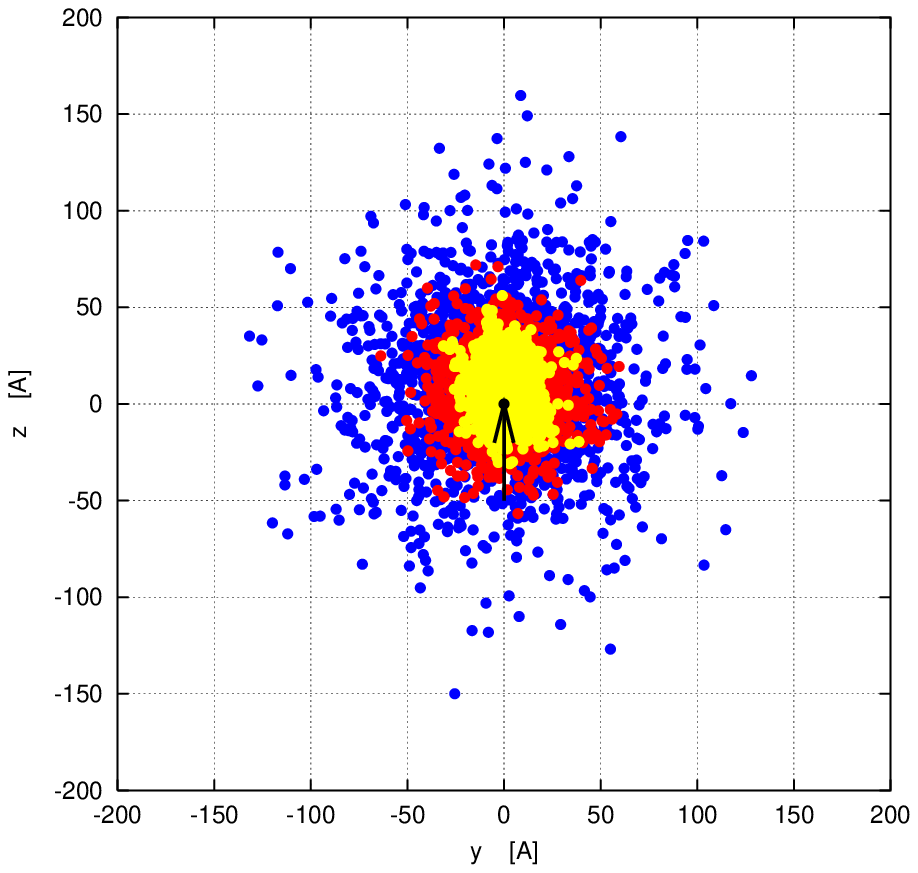}\\
\end{center}
\caption{{\footnotesize Transverse sections of the electron cloud obtained from the TPP-2 model when no energy transfer to the lattice is allowed in planes: {\bf (a)} XY; {\bf (b)} XZ; {\bf (c)} YZ. The thickness of the section is $40$ \AA . Velocity of the impact electron at $t=0$ fs points towards the Z-axis. 
}}
\label{f8}
\end{figure}
%

Fig.\ \ref{f1} presents the band structure of diamond at $T=300$ K, as derived from \cite{diamond1}. The zero of the energy scale lies in the middle of the band gap. The position of the bottom of the conduction band is $E_C=E_G/2$, and the band gap equals $E_G=5.46$ eV at $T=300$ K. 
The width of the valence (V) band is estimated to be $\Delta E_V\sim 23$ eV \cite{diamond2}.  
The dispersion relations for electrons near the bottom of the conduction band and for holes near the top of the valence band are given by the quadratic approximation \cite{l6}. In order to keep consistency with the Lindhard dielectric formulation, we assume henceforth the effective mass of the electrons in the conduction band to be equal to the mass of a free electron in vacuum, $m_e$. For simplicity, we also assume that the mass of the hole is direction-independent, and equals to the effective mass, $m_V=0.8 m_e$, where $m_V^3$ is the determinant of the valence band mass tensor \cite{l6,diamond3}. In contrast with the free-electron-gas model of the insulator band \cite{ziaja}, the kinetic energies of free electrons are measured with respect to the bottom of the conduction band (see Fig.\ \ref{f1}). Following Ref.\ \cite{l6}, the number of electrons or holes in an infinitesimal energy interval $(E,E+dE)$ for an insulator is:
\eq
dn(T)=dE \left\{ g_C(E){\bar \Theta(E>E_C)} + g_V(E){\bar \Theta(E<E_V)}\right\}
f(E,\mu,T),
\label{dn}
\eqx
where $g_{C,V}(E)$ corresponds to the density of levels in conduction or valence band,
\eq 
g_{C,V}(E)=\sqrt{2\mid E-E_{C,V}\mid}{{m_{C,V}^{3/2}} \over
{\hbar^3\pi^2}},
\label{gcv}
\eqx
and $f(E,\mu,T)$ describes the Fermi-Dirac distribution of electrons, depending on the chemical potential $\mu$. The function $\bar \Theta$ is the step function.

\subsection*{Electron trajectories:}

The trajectory of electrons inside a diamond bulk was simulated in the following way: the probability that a free electron of kinetic energy $E$ travels a distance $x$ in a solid without being scattered decreases exponentially, $P(E,x)=exp(-x/\lam_T(E))/\lam_T(E)$, where $\lam_T(E)$ is the total electron mean free path:
\eq
\frac{1}{\lam_T(E)}=\frac{1}{\lam_{el}(E)}+\frac{1}{\lam_{in}(E)}.
\label{llam}
\eqx

For a given (probability) distribution $f(y)$ of $y$, we define a normalized probability distribution, $P(y)$, in the form:
\eq
P(y)={ {\int_{y_{min}}^y\,f(y^{\prime})dy^{\prime}}
\over
{\int_{y_{min}}^{y_{max}}\,f(y^{\prime})dy^{\prime}}},
\label{mc}
\eqx
where $y_{min}$ and $y_{max}$ limit the allowed values of $y$, $y_{min}<y<y_{max}$. It may be easily checked that $0\leq P(y)\leq 1$.
In the MC simulation $y$ is then chosen to fulfill the relation: $r=P(y)$, 
where $r$ is a random number, $0<r<1$.

The path length between two collisions may then be obtained from the normalized probability distribution, and it reads $x=-\lam_T(E)\ln(r)$, where $r$ is a random number, $0<r<1$.

After the electron has travelled a distance x, either elastic or inelastic collisions occur as a stochastic process (probability of collision $\approx\frac{\sigma_{el(in)}}{\sigma_{el}+\sigma_{in}}$). In elastic collisions, the electron preserves its energy, but it is scattered at an angle $\Theta_{el}$. The value of $\Theta_{el}$ is obtained from the normalized probability distribution (\ref{mc}) obtained from the differential elastic cross section $\seld$. The momentum transfer to the lattice is neglected at this point. For an inelastic collision the situation gets more complicated. The scattered electron loses part of its energy $\hbar\om$ and changes its momentum by $\hbar q$. From the value of $\hbar q$ one can estimate the inelastic scattering angle $\Theta_{in}$. MC probabilities $P(E,\om,q)$ for this process are obtained from the doubly differential cross section $\displaystyle \frac{d^2\sigma_{in}}{dq\,d\om}$.

The energy $\hbar \om$ and momentum $\hbar q$ losses can then be transferred to excite a secondary electron from the valence band into the conduction band. The excited electron leaves a hole with an opposite momentum in the valence band. However, the information whether a secondary electron excitation took place, or not, is not contained implicitly in the inelastic cross section. In order to estimate the number of secondary emissions, we consider two limiting cases
here: 

{\bf Limiting case 1}: Let us introduce a probability, $P(E_0)$, to excite an electron of initial energy $E_0$ from the valence band. The probability is derived from the carrier distribution (\ref{dn}), using (\ref{mc}). In the first case we assume that the probabilities $P(E,\om,q)$ and $P(E_0)$ describe two independent events: (a) a scattering of the primary electron of energy $E$ with energy loss $\hbar \om$ and momentum transfer $\hbar q$, and (b) a possible excitation of an electron of the initial energy $E_0$ from the valence band into the conduction band, if $E_0+\hbar \om$ is greater than $E_C$. If, however, $E_0+\hbar\om < E_C$, the excitation cannot take place, and the energy loss of the primary electron is transferred to the crystal lattice. 

{\bf Limiting case 2}: Here we do not allow energy transfer to the lattice. We assume that the energy lost by the primary electron is given to a secondary electron (and a hole). The electron is then removed from the valence band into the conduction band without any energy transfer to the lattice. Events (a) and (b) above are thus no longer independent, and a joint probability $P(E,\om,q,E_0)$ replaces the product $P(E,\om,q)P(E_0)$. This approach gives the upper limit for the number of secondary electrons released.

In the first limiting case the momentum conservation imposed on the secondary electron emission process, reduces the number of emitted secondary electrons. The electron excited from valence band has an initial kinetic energy of $E_{k,sec}^e=\frac{m_V}{m_e}(E_V-E_0)$ obtained from the dispersion relation in the valence band. This corresponds to the kinetic energy of the hole, $E_{k,sec}^h=E_V-E_0$ in the valence band (see Fig.\ \ref{f11}a). The energy $E_V$, $E_V=E_C-E_G$, is the energy at the top of the V band. The final kinetic energy of the excited electron is then $E_{k,sec}^{\prime}=E_0+\hbar\om-E_C$. 
The incoming electron transfers a part of its momentum, $\hbar {\bf q}$, to the secondary electron and the hole. Momentum conservation requires that ${\bf p}_{sec}=-{\bf p}_h$ and ${\bf p}_{sec}^{\prime}= -{\bf p}_h-\hbar {\bf q}$ (see Fig.\ \ref{f11}b), where ${\bf p}_{sec}$ and ${\bf p}_{sec}^{\prime}$ are the initial and final momenta of the secondary electron, and ${\bf p}_h$ is the momentum of the hole. The momenta of electron have magnitudes $\mid {\bf p}_{sec}\mid \sim \sqrt{2m_e\,E_{k,sec}}$, $\mid {\bf p}_{sec}^{\prime}\mid \sim \sqrt{2m_e\,E_{k,sec}^{\prime}}$, and their orientations are chosen to fulfill the conservation of the momentum, which requires: 
{\footnotesize $| \mid {\bf p}_{sec}\mid -\mid {\bf p}_{sec}^{\prime}\mid | < \hbar \mid {\bf q} \mid < \mid {\bf p}_{sec}\mid + \mid {\bf p}_{sec}^{\prime}\mid$ }. This is not always possible at independently chosen $E_0$'s and $(\om, q)$ pairs. 

For both secondary emission schemes, we write the energy conservation law for the cloud of electrons in the form~:
\eq
E_{kinet,T}^e+E_{kinet,T}^h=E -N_{ion}\cdot E_G - E_{loss},
\label{econs}
\eqx
where $E_{kinet,T}^e$ is the total kinetic energy of electrons in the conduction band, $E_{kinet,T}^h$ is the total kinetic energy of holes in the valence band, $E$ denotes the impact energy of the primary electron, $N_{ion}$ denotes the number of secondary electrons (holes) released, and $E_{loss}$ is the energy lost in inelastic collisions without secondary electron excitations. The ionization of the medium continues, until the energies of all excited electrons, including the primary electron, fall below $E_{elast}$. After that point, only elastic scatterings occur. 

\subsection*{Numerical results}


A set of 2000 MC simulations in diamond was performed. Each cascade was initiated by a single electron. In these simulations, the kinetic energy of the primary electron, measured in respect to the bottom of the conduction band was fixed at $E=250$ eV. The starting position of the primary electron at $t=0$ fs was at the point {\bf x}=(0,0,0) of an arbitrarily chosen coordinate system. Afterwards, the electron cloud was created via secondary electron emissions. Motions of the holes were neglected.

The value of the elastic and inelastic cross sections depends on the
energy of the electron, and this is changing with time. The elastic and
inelastic mean free paths of low energy electrons were computed as
described in \cite{ziaja}. Within the energy regime considered here, inelastic
collisions were rare compared to elastic interactions, and most of them
happened within the first $10$ fs following primary electron impact. Within
the first $100$ fs, there were about $10000$-$35000$ elastic interactions
and about $10$-$50$ inelastic interactions in an individual cascade simulation, from which $5$-$13$ resulted in the
ionization of the sample. The data show that elastic interactions represent 
the dominant mechanism for electron propagation in the system.
 
{\bf Spatio-temporal evolution} of the cascade was analysed through {\bf (a)} the number of secondary ionizations, $N_{ion}(t)$, {\bf (b)} the electron range, $r_{max}(t)$, i.\ e.\ the distance of the most distant electron from the position of the primary electron emission, {\bf x}=(0,0,0), {\bf (c)} the kinetic temperature of the free electron gas $kT(t)$. It should be noted that the kinetic temperature is a non-equilibrium parameter. Quantities (a)-(c) were averaged over a number of cascades. Figure \ref{f3} shows the results.

The average number of ionization events during the first femtosecond was estimated to be $4-6$ based on Ashley's model and $5-8$ based on the TPP-2 model. This is due to the fact that Ashley's inelastic mean free path, considered as a function of kinetic energy in the conduction band, is larger than the corresponding TPP-2 inelastic mean free path at all energies, and less ionizations occur within the same time interval. The number of secondary ionizations increased with time, and saturated within about $\leq100$ fs with a total of $5-10$ electrons (Ashley), or $6-13$ (TPP-2).   

It should be stressed that the ionization rate is very sensitive to the secondary electron emission scheme used in the calculations. However, all predicted ionization rates lie within the limiting values predicted by experiment and other phenomenological models \cite{secon1,secon2,secon3,secon4,secon5}. The lower limit is given by the secondary electron emission yield, $\delta$, defined as the ratio of the intensity of secondary electrons emitted from the surface to the incident beam intensity. Its experimental value at $E\sim 250$ eV is $\delta\leq 3$ \cite{secon1,secon4}. This means that at least three electrons were created in the cascade. The upper limit of ionization rate, $N_{ion}^{max}$ is estimated \cite{secon3,secon4,secon5} as $N_{ion}^{max}= E/ E_{e-h}$, where $E$ is the energy of the primary electron, and $E_{e-h}$ is the average energy needed to create an electron-hole pair in an insulator. For diamond $E_{e-h}\sim 17$ eV \cite{secon3}, so $N_{ion}^{max}\sim 15$. Our model gives: without energy transfer to the lattice: $10$-$13$ secondary electrons, with energy transfer: $5$-$6$ secondary electrons.

The average electron range after $90$ fs was around $110-130$ \AA. The electron range predicted by Ashley's model was about 24\% larger than in the TPP-2 model at $90$ fs. Both models predicted almost linear growth of $r_{max}$ as a function of time after $\sim40$ fs. This is due to the fact, that after this time low energy electrons are dominant in the sample, and, as a consequence, 
the dynamics of the cloud is dominated by isotropic elastic scatterings. It should be stressed that for low energy electrons ($E\leq 10$ eV)  phonon coupling becomes important in insulators. Here, we 
did not take phonon exchanges into account. 

The temperature of the electrons drops as the cascade evolves. The temperature characteristics of the electron cloud in Ashley's and the TPP-2 model are similar. In both models they are quite insensitive to the different limiting cases of secondary electron emission. The curves show rapid decrease within $1$ fs after the primary electron emission. This is due to the fact that most ionizations occured during the first femtosecond, and subsequent expansion cools the system effectively. The rate of cooling drops later. Temperature data were used to check that in all simulations the energy was conserved in the system.

{\bf Energy distribution of secondary electrons}. The positions and velocities of electrons recorded at times, $t=1,\,10,\,90$ fs were collected from all cascades generated and put into one file. Using these data, histograms of energy distributions were obtained (Fig.\ \ref{f4}), $N(E)/N$, at these time points ( $N(E)=\sum_{i=1}^{2000}\,N_i(E)$, and $N_i(E)$ was the number of electrons found in the energy interval $(E,E+\Delta E)$ for the ith cascade ). The distributions were normalized to the total number of electrons, $N=\sum_{E}\,N(E)$. 

As expected, the energy distribution of secondary electrons shows that the number of low energy electrons increases with time. At $10$ fs around $30-80$\% of electrons have energies lower than $10$ eV. At $90$ fs there are mainly low energy electrons (more than $50$\% of electrons of energy, $E<10$ eV) in the sample. Average total energy losses to the lattice, $E_{loss}$, are $\sim 106$ eV in Ashley's model and $\sim 95$ eV in Tanuma's model within $100$ fs for the first secondary electron emission scheme, i.e. when energy transfer to the lattice is allowed.

{\bf Local electron density and the Debye length}. 
Using the cumulative results from all cascades, we calculated the
average local electron density, $\rho(r)$, and the local electron temperature in spherical shells of the width $1$ \AA, centered at the center of mass of the cumulative electron cloud. Using these data, we calculated the Debye 
length, $\lam_D(r)$ \cite{plasma1}, in each shell. The results recorded at 
times, $t=1,\,10,\,90$ fs, are shown in Figs.\ \ref{fig1}-\ref{fig4}.
The results show that the Debye length was longer than 
the radial size of the electron cloud. This was also the case at the highest density peak. There is an approximate exponential relation between the Debye length, $\lam_D(r)$, and the radial size of the electron cloud, $r$, $\ln \lam_D(r)=a\,r +b$, which in this case implies that $\lam_D(r) > r$.  Therefore the long-range Coulomb field is not shielded within this cloud, and the electron gas generated does not represent a plasma in a single impact cascade triggered by an electron of $E\sim 250$ eV energy.
This is important as it justifies the independent-electron approximation used in the model. Moreover, crude estimations of the ionization fraction show that the ionization level needed to create an Auger electron plasma in diamond will be reached with a dose of $\sim 2\times 10^5$ impact X-ray photons per \AA$^2$ if these photons arrive before the cascade electrons recombine.

{\bf Spatial distribution of secondary electrons}. In order to analyse the spatial distribution of secondary electrons, positions of all electrons generated in $2000$ cascading processes were collected (at times, $t=1,\,10,\,90$ fs). The $40$ \AA $\,\,$ wide cross sections of the collective clouds are shown in Figs.\ \ref{f5}-\ref{f8}. They were obtained for different optical models and different secondary electron emission schemes. In all models, most 
of the electrons lay within a sphere of radius $\leq 50$ \AA $\,$ at $1$ fs, $\leq 70$ \AA $\,$ at $10$ fs and $\leq 150$ \AA $\,$ at $90$ fs with respect to the origin of the cascade.
It was found that the center of mass of the cloud was moved $\sim 7$ \AA $\,$ from the starting point in the direction of the electron impact with $250$ eV
electrons. The shifts in other directions were $\leq 1$ \AA. 

The sphericity tensor, 
$\displaystyle S^{ab}={{\sum_{i=1}^{N}\,r_i^a r_i^b}\over
{\sum_{i=1}^{N}\,r_i^2}}$, 
calculated for the collective cloud was almost diagonal. Non-diagonal elements were $\sim 10-1000$ smaller than the diagonal ones. The sphericity tensor was diagonalized, and the sphericity parameter, $S$, was obtained from its eigenvalues: $S=3/2\,(\lambda_2+\lambda_3)$, where $\lambda_1 \geq \lambda_2 \geq \lambda_3$. For all emission schemes $S$ lay within the interval $0.83-0.89$ at $1$ fs, $0.95-0.97$ at $10$ fs and $0.98-0.99$ at $90$ fs. This implies that the spatial distribution of the collective cloud was practically isotropic, since $S\rightarrow 1$.

A small anisotropy appeared only at $1$ fs. It manifested distinctly in the cloud generated in the TPP-2 model when no energy loss to the lattice was allowed (cf.\ Fig.\ \ref{f8}). The distribution was cylindrically symmetric along the primary impact
vector but it shrank along the direction of the primary impact. This was due to the fact that at that time scale the primary electron was still much faster than the secondary electrons (cf. Fig.\ \ref{f4}a). At later times scales the sample got dominated by low energy electrons which scattered more and more isotropically. This was reflected in the progressing isotropy in the distribution. 


\subsection*{Conclusions} 

The mechanism of ionization in solids by low energy electrons 
($\sim 10$-$500$ eV) is not trivial. When the electron wavelength is 
comparable to distances between atoms in a solid, the Born approximation breaks down and a quantum mechanical description of the electron-atom interactions is necessary. Such a model has recently been published \cite{ziaja} on the basis
of a free-electron gas approximation for the electronic structure of the
solid. While this approach is appropriate for the description of metallic
substances, it is inadequate for insulators. Insulators have a closed
valence band and the band gap between the valence band and the conduction
band is large. Here we describe simulations where the band structure of
the insulator (diamond) was explicitly included. Electron impact cascades
were initiated by a single electron of $250$ eV, corresponding to the Auger
energy of carbon. The results of Monte Carlo simulations show the
spatio-temporal evolution of the ionization cascades with the number of
secondary electrons emitted, the range of emitted electrons, their kinetic
temperature, energy distribution and the plasma characteristics in terms
of the Debye length. The following conclusions can be drawn from this
study:

- Ionization rates

The average number of ionization events during the first femtosecond was
between $4-8$ depending on the secondary electron emission scheme used in
the calculations. Saturation was reached within $\leq100$ fs with a
total of $5-13$ in the various models. These ionization rates are lower
than rates obtained earlier with the Fermi electron gas approximation
\cite{ziaja}, and agree better with experiments and other predictions
\cite{secon1,secon2,secon3,secon4,secon5}. At time scales much longer than
$10$ fs, very low energy electrons dominate the sample, and a model specific
for this energy regime should be worked out. 

- Energy distribution

The average temperature of the electrons dropped rapidly within the first
femtosecond following the primary emission. This is due to the fact that a
large part of all ionizations occurred during the first femtosecond, and
this cooled the system effectively. As expected, the number of low energy
electrons increased with time, and very low energy electrons ($E<10$ eV)
became dominant in the sample after about $10$ fs. As a consequence, elastic interactions were frequent in the sample. 
When energy transfer to the lattice was allowed, cooling was faster, with an average total energy loss to the lattice of around $100$ eV within $100$ fs
in an individual cascade simulation.

- Spatial distribution

The spatial distribution of secondary electrons is anisotropic towards the
primary impact at $1$ fs. At longer time scales, as the system cools down
and energy is distributed more and more equally among electrons, the
spatial distribution of the electrons becomes isotropic. Phonon coupling
and inelastic interactions between very low energy electrons ($E<10$ eV) and matter were neglected in the model. These effects may influence the
transportation of very low energy electrons within the sample.

- Debye length and Coulomb shielding

The Debye length increases exponentially with the radial size of the
electron cloud. The long-range Coulomb field is not shielded within this cloud, and the electron gas generated does not represent a plasma in a single impact
cascade triggered by an electron of $E\sim 250$ eV energy. This is
important as it justifies the independent-electron approximation used in
the model. Moreover, an analysis of the ionization fraction shows that ionization level needed to create an Auger electron plasma in diamond will be reached with a dose of $\sim 2\times 10^5$ impact X-ray photons per \AA$^2$ if these photons arrive before the cascade electrons recombine. 

- Implications

The results can be used to estimate damage by low energy electrons in
diamond and other carbon-based covalent compounds. The Monte-Carlo code may be
adopted to simulate multiionization phenomena in systems, ranging from the
explosion of atomic clusters to the formation of plasmas. The model could
also be used to estimate ionization rates and the spatio-temporal
characteristics of secondary electron cascades in biological
substances. Based on numerical figures for diamond, the maximal diameter
of the secondary electron cloud reaches about $100$  \AA ngstrom within the
first femtosecond with about $4$-$6$ ionization events in this period. The
diameter of this cloud is $2$-$3$ times larger at than the diameter of a
simple protein molecule. At $10$ fs, ionization is almost complete with
an average of $5$-$11$ electrons released per cascade. At this point the
maximal diameter of the cascade is about $150$ \AA ngstrom, i.e. the size of a
large multienzyme complex. At $100$ femtosecond, the diameter of the
secondary electron cloud reaches the size of a small virus particle
($300$-$400$ \AA ngstrom). The overall number of ionization events does not
change appreciably from this point onwards as the
electrons have cooled down and interact elastically with the sample. Note that
the density of proteins is about $3$ times smaller than that of diamond, 
and so the total number of interactions and ionization rates are expected 
to be lower in proteins than in diamond. The
figures above overestimate ionization within the volumes of small
biological samples at these time points. Further studies will be necessary
to quantitate expected differences. 

One should also point out that there are certain practical applications
for diamond in which an understanding of the damage mechanisms is
important. For instance, diamond and other wide band gap materials are
considered as potential cold electron emitters. Due to its low or negative
electron affinity, diamond irradiated by electrons shows a high secondary
electron emission yield \cite{secon1,secon2}. This offers the possibility
for applying diamond for signal amplification, e.\ g.\ in scanning
electron microscopy. The high radiation resistivity of diamond makes it an
ideal candidate to work as a semiconductor in high-radiation environments
\cite{camp}, as, for instance, in particle detectors. 


\section*{Acknowledgments} 

We are grateful to Gyula Faigel, Zoltan Jurek, Michel A. van Hove, Richard W. Lee, Richard London, Leszek Motyka and Edgar Weckert for discussions. This research has been supported in part by the Polish Committee for Scientific Research with grant No.\ 2 P03B 05119, the EU-BIOTECH Programme and the Swedish Research Councils. A.\ S.\ was supported by a STINT distinguished guest professorship. B.\ Z.\ was supported by the Wenner-Gren Foundations.


\end{document}